\documentclass[11pt,english,a4paper]{article}
\usepackage[utf8]{inputenc}
\usepackage[longnamesfirst]{natbib}

\usepackage[top=32truemm,bottom=32truemm,left=25truemm,right=25truemm]{geometry} 

\usepackage{amsthm, amssymb, amsmath, mathtools, physics, enumitem}
\usepackage[title]{appendix}
\usepackage{bm}
\usepackage{xcolor}

\usepackage{booktabs}
\usepackage{threeparttable}
\usepackage{subcaption}
\usepackage{hyperref}
\usepackage{mathtools}
\mathtoolsset{showonlyrefs=true}

\usepackage[mathlines]{lineno}
\theoremstyle{definition}
\newtheorem{theorem}{Theorem}
\newtheorem{lem}{Lemma}
\newtheorem{definition}{Definition}
\newtheorem{proposition}{Proposition}
\newtheorem{corollary}{Corollary}

\newtheorem{claim}{Claim}

\usepackage{tikz}
\usetikzlibrary{intersections, calc, arrows.meta}
\makeatletter

\@addtoreset{figure}{section}
\@addtoreset{table}{section}

\makeatother

\title{Collective decisions under uncertainty:\\ efficiency, ex-ante fairness, and normalization%
\thanks{The authors are grateful to three anonymous referees for their helpful comments and suggestions.
The authors also thank Tsuyoshi Adachi, Susumu Cato, Itzhak Gilboa, Takashi Hayashi, Noriaki Kiguchi, Daiki Kishishita, Kaname Miyagishima, Nozomu Muto, Jawwad Noor, Satoshi Nakada, Hendrik Rommeswinkel, Koichi Tadenuma, Norio Takeoka, Tsubasa Yamashita, and Shohei Yanagita for their insightful comments and discussions. 
This paper has been presented at the Japanese Economic Association 2025 Spring Meeting (Chukyo University), the World Congress of Econometric Society 2025 (Seoul), Tokyo Social Choice Theory Workshop (the University of Tokyo), and seminars held at Aoyama Gakuin University and Tokyo University of Science. 
This research is financially supported by KAKENHI (Nos.~25KJ2146 \& 25KJ1298) and Waseda Institute of Political Economy (WINPEC).}}
\author{Leo Kurata\thanks{Graduate School of Economics, Waseda University, 1-6-1 Nishiwaseda, Shinjuku-ku, Tokyo 169-8050, Japan. E-mail: \url{leo.kurata.ac@gmail.com}}\hspace{1mm} and Kensei Nakamura\thanks{Graduate School of Economics, Hitotsubashi University, 2-1, Naka, Kunitachi, Tokyo 186-8601, Japan.  E-mail: \url{kensei.nakamura.econ@gmail.com}}}
\date{Last update: \today}

\begin{document}

\maketitle
\vspace{-5mm}

\begin{abstract}
This paper studies preference aggregation under uncertainty in the multi-profile framework and characterizes a new class of aggregation rules that can address classical concerns about \citeauthor{harsanyi1955}'s (\citeyear{harsanyi1955}) utilitarian rules. 
Our aggregation rules, which we call \textit{relative fair aggregation rules}, are grounded in three key ideas: utilitarianism, egalitarianism, and the 0--1 normalization of individual utilities. 
These rules are parameterized by a set of weight vectors over individuals and evaluate each ambiguous alternative by taking the minimum weighted sum of 0--1 normalized utility levels over the weight set. 
For the characterization, we propose two novel axioms, \textit{weak preference for mixing} and \textit{restricted certainty independence}, with normative justifications.
Additional results clarify how these axioms capture the utilitarian and egalitarian attitudes of the rules.
\\
\vspace{3mm}

\end{abstract}

\newpage

\section{Introduction}\label{sec1}
\citet{harsanyi1955} studied how to construct social preferences based on individual values under risk.\footnote{
Risk refers to situations in which the outcomes of an action are not known in advance, but the probabilities associated with these outcomes are given or objectively specified.
}
Harsanyi's aggregation theorem states that if individuals and the social planner have expected utility (EU) preferences, only weighted utilitarian aggregation rules satisfy the Pareto principle. 
Since this result relies only on two widely accepted conditions---EU rationality and the Pareto principle---it has been regarded as a primary foundation for utilitarianism under risk.  

Despite its appeal, this theorem has long drawn criticism from many scholars, especially normative economists and philosophers. 
We summarize the major issues with Harsanyi's theorem and utilitarianism itself:
\begin{enumerate}[label=(\roman*)]
\setlength{\itemsep}{3pt}
\setlength{\parskip}{0pt}
    \item \textit{Indeterminacy of weights.} 
    For any social preference relation represented by a weighted sum of individual von Neumann--Morgenstern (vNM) functions, we can obtain another representation of the relation with a different weight vector by rescaling each individual vNM function appropriately.%
    \footnote{These adjustments are permitted by the uniqueness theorem for vNM functions.}
    Consequently, the weights assigned to individuals are not uniquely determined, which undermines ethical interpretations of Harsanyi's theorem (cf.\ \citet{sen1976welfare}, \citet{weymark1991}). 
    
    \item \textit{Distributive justice.} Utilitarianism is indifferent to welfare distributions as long as the total welfare is unchanged (e.g., \citet{rawls1971}). 
    Thus, utilitarianism sometimes leads to unacceptable consequences from an equality perspective, such as extreme situations in which one individual possesses all resources while everyone else has none. Furthermore, utilitarianism under risk cannot account for equality of opportunity (cf.\ \citet{diamond}).
    \item \textit{Uncertain situations.} In practice---for example, when assessing environmental or monetary policies---decisions are often made under uncertainty (i.e., in situations where agents do not know the ``true'' probability distribution). 
    As \citet{mongin1995consistent} pointed out, the analogue of Harsanyi's theorem no longer holds in this context: 
    Under the assumption that agents and society are EU maximizers, the Pareto principle entails dictatorship unless individual beliefs are identical.
    This suggests that once we consider broader situations, Harsanyi's justification for utilitarianism loses its validity. 
\end{enumerate}

The objective of this paper is to address these issues by proposing a new class of collective decision-making rules under uncertainty. 
Specifically, we analyze \citeauthor{sprumont2018belief}'s (\citeyear{sprumont2018belief,sprumont2019relative}) framework, in which feasible outcomes and preference profiles may vary depending on the situation that society faces.
Given a set of feasible deterministic outcomes, uncertain prospects are formalized as acts \`a la \citet{savage1}, that is, as functions from states to outcomes.
A \textit{problem} is a social situation formalized as a pair consisting of a set of feasible outcomes and a profile of subjective EU preferences over acts. 
Within this framework, we axiomatically study aggregation rules, defined as mappings that associate each problem with a social preference relation, from a normative perspective.
 
Our aggregation rules are grounded in three key ideas: utilitarianism,  egalitarianism, and the 0--1 normalization of individual utilities. 
These ideas are reflected in the structure of the rules.
Specifically, each rule in the class is associated with a set of weight vectors, say $\mathcal{M}$, and evaluates an act based on individual ex-ante evaluations through the following steps:
(1) Given a problem, normalize each individual's utilities so that the maximum and minimum of the range are $1$ and $0$, respectively;
(2) compute the weighted sum of these normalized expected utilities for each weight vector $\mu\in \mathcal{M}$;
and (3) adopt the minimum weighted sum as the evaluation of the act.
The 0--1 normalization in the first step provides a natural basis for the interpersonal comparison of utilities.
Fixing the range of utility levels eliminates the arbitrariness in the choice of utility representations, which in turn offers a clear ethical interpretation of the weights.
In the second and third steps, the evaluation proceeds as in the maxmin EU model of \citet{gilboa1989maxmin}:
The rule tends to assign higher weights to individuals who are disadvantaged with respect to normalized utility levels.
Hence, the last two steps allow for egalitarian considerations while maintaining a weighted-utilitarian approach.

The weight set can be interpreted, in general, as parameterizing the extent to which the social planner cares about those who are worse off in terms of normalized utility levels.
If, on the one hand, the weight set is a singleton, the rule reduces to utilitarianism with the 0--1 normalization. 
On the other hand, if the weight set coincides with the entire set of weight vectors over individuals, the functional form
corresponds to the maximin rule (cf.\ \citet{rawls1971}) with the 0--1 normalization.
The former rules constitute what is known as ``relative utilitarianism'' and have been widely studied in the literature; see, for example, \citet{dhillon1999relative} and, in the present framework, \citet{sprumont2019relative}. 
Since our aggregation rules incorporate a notion of fairness into relative utilitarianism, we refer to them as \textbf{relative fair aggregation rules}.
By allowing the weight set to lie between the two extremes, our aggregation rules accommodate various attitudes toward efficiency and equity.

Our main theorem (Theorem~\ref{thm_relativefair}) characterizes the relative fair aggregation rules using two inter-problem and two intra-problem axioms, along with the basic axioms, namely the Paretian condition and continuity.
The inter-problem axioms mainly serve to establish the 0--1 normalization of individual utilities.
The axiom called \textit{independence of redundant outcomes} requires the invariance of social evaluation under ``redundant expansions'' of feasible sets. 
Specifically, it postulates that if each added outcome is redundant, in the sense that there exists an originally feasible act that yields the same utility to all individuals, the social ranking of acts feasible in the original problem remains unchanged.%
\footnote{
A similar axiom was first introduced by \citet{dhillon1999relative} under risk. 
In the present framework,  \citet{sprumont2019relative} also discussed \textit{independence of redundant outcomes} but did not impose it in the characterization. Section~\ref{sec:4-2} provides further details.}
We show that \textit{independence of redundant outcomes} characterizes the 0--1 normalization 
under the basic axioms and the other inter-problem axiom called \textit{belief irrelevance}, which posits that the evaluation of unambiguous outcomes is independent of individual beliefs. 
That is, these axioms yield a social welfare function over the 0--1 normalized utility vectors (Lemma~\ref{lem:basic}). 

Meanwhile, the two intra-problem axioms restrict how the social planner evaluates normalized utility vectors. 
The first intra-problem axiom is \textit{weak preference for mixing}, which
requires that mixing outcomes via a fair coin toss be socially desirable. 
More specifically, it demands that any act that all individuals believe yields two outcomes with equal probability be not worse than at least one of the constant acts in which either outcome occurs with certainty. 
Since the former act is more equitable than the constant acts from the ex-ante perspective, this axiom embodies a fairness notion in aggregation rules (see Section~\ref{subsec:4-3} for details). 
Indeed, together with the axioms in Lemma~\ref{lem:basic}, \textit{weak preference for mixing} characterizes the quasiconcavity of the social welfare function over the normalized vectors (Lemma~\ref{lem:quasiconcav}). 

The other intra-problem axiom concerns the consistency of evaluations within a problem, which has been widely studied in decision theory under the label of the independence axiom.
Since we study the Savage setting, objective mixture operations over alternatives are not available, unlike in the other popular setting of \citet{anscombe1963definition}. 
Nevertheless, we show that by utilizing subsets of the state space to which all individuals assign the same probabilities, one can define a \textit{pseudo-mixed act} that 
each individual perceives as if it were constructed by randomizing two acts (Proposition~\ref{prop:pmact}).  
Using this mixture operation, we introduce \textit{restricted certainty independence} as a novel axiom.
It states that whenever vNM functions are essentially identical across individuals, the social preference relation between two acts is preserved even after those acts are pseudo-mixed with some outcome in a common proportion, akin to the certainty independence axiom of \citet{gilboa1989maxmin}. 
Restricting the scope of the axiom to problems in each of which all individuals share a common vNM function is necessary to avoid imposing consistency conditions in seemingly unsuitable situations where, for instance, the mixed outcome is desirable for some individuals but unacceptable for others (see the argument in Section~\ref{subsec:4-4}).
Together with the axioms in Lemma~\ref{lem:basic}, \textit{restricted certainty independence} yields the weak linearity (i.e., homogeneity and constant-additivity) of the social welfare function over the normalized utility vectors (Lemma~\ref{lem:4addCI}). 
That is, \textit{restricted certainty independence} reflects the utilitarian aspect of the relative fair aggregation rules.

Building on the main theorem, we axiomatize the two extreme cases as well.
By strengthening \textit{restricted certainty independence} and abandoning \textit{weak preference for mixing}, we obtain the relative utilitarian rules (Theorem~\ref{thm:rel-util}). 
In contrast, under the impartiality axiom,  abandoning the former and strengthening the latter yield the maximin rule with the 0--1 normalization (Theorem~\ref{thm_relmaxmin_mix}).
These results highlight how the two intra-problem axioms capture two core concepts of our aggregation rules: utilitarianism and egalitarianism.

Furthermore, we show based on Lemma \ref{lem:basic} that the second special case is characterized by another axiom concerning social views toward uncertainty (Theorem~\ref{thm_ambavo_maxmin}). 
This condition requires that for any problem where vNM functions are essentially identical across individuals, 
the social planner should be more uncertainty-averse than individuals in the sense of \citet{ghirardato2002ambiguity}. 
This axiom itself specifies only the social planner's comparative uncertainty attitude, 
but when combined with the other axioms, it ultimately leads to the strongest concern for the relatively disadvantaged individual.

This paper is organized as follows: 
Section~\ref{sec2} presents the formal setup. 
Section~\ref{sec3} formalizes the relative fair aggregation rules, and
Section~\ref{sec4} characterizes the rules step by step.
Section~\ref{sec5} provides characterizations of the two extremes. 
Finally, Section~\ref{sec6} discusses our results in relation to the literature. 
All proofs appear in Appendix~\ref{app}, while Appendix~\ref{app_general} provides a further result.

\section{Setup} \label{sec2}

Our framework is based on \citet{sprumont2018belief,sprumont2019relative}.
Let $(\Omega, \Sigma)$ be a measurable space, where $\Omega$ is a set of states of nature and $\Sigma$ is a $\sigma$-algebra of subsets of $\Omega$. 
Each element of $\Sigma$ is called an \textit{event}.
The infinite set of potentially feasible outcomes is denoted by $\mathbb{X}$. We assume that a set $X$ of \textit{feasible outcomes} is a finite subset of $\mathbb{X}$ and that its cardinality is at least two.
The collection of sets of feasible outcomes is denoted by $\mathcal{X}$, that is, $\mathcal{X}$ is the set of finite subsets $X$ of $\mathbb{X}$ such that $|X| \geq 2$.

For any $X\in \mathcal{X}$, an \textit{$X$-valued act} is a function $f: \Omega \rightarrow X$. 
When we do not need to mention the range, we simply call it an act. 
The set of $X$-valued acts is denoted by $F_X$. 
With a slight abuse of notation, we identify an outcome $x\in X$ with the constant act $f \in F_X$ such that for all $\omega\in \Omega$, $f(\omega) = x$.
For any $X\in\mathcal{X}$, any $x, y \in X$, and any event $E\subset \Omega$, let $xEy \in F_X$ be the act such that $(xEy) (\omega) = x $ for all $\omega \in E$ and $(xEy) (\omega) = y $ for all  $\omega \notin E$. 

Let $N = \{ 1, 2, \ldots, n \}$ be a fixed set of individuals, with $n\geq 2$. 
Given $X\in \mathcal{X}$, each individual $i$ has a complete and transitive preference relation $R_i$ over $F_X$.%
\footnote{A binary relation $R$ over $F_X$ is complete if, for any $f,g \in F_X$, $f R g$ or $g R f$. 
Also, $R$ is transitive if, for any $f,g,h \in F_X$, $f R g$ and $g R h$ imply $f R h$.} 
For $f,g\in F_X$, when we write $f R_i g$, it means that individual $i$ regards $f$ as at least as desirable as $g$. 
The symmetric and asymmetric parts of $R_i$ are denoted by $I_i$ and $P_i$, respectively.
For $X \in \mathcal{X}$, the set of complete and transitive binary relations over $F_X$ is denoted by $\mathcal{R}(X)$. 
Let $\mathcal{R} = \bigcup_{X\in \mathcal{X}}\mathcal{R} (X)$. 

We assume that for each $i\in N$, $R_i$ follows \citeauthor{savage1}'s (\citeyear{savage1}) \textit{subjective expected utility (SEU)} theory.
That is, given $X\in \mathcal{X}$, there exist a nonconstant function $u_i : X\rightarrow \mathbb{R}$ and a countably additive, nonatomic probability measure $p_i$ on $\Sigma$ such that for all $f, g\in F_X$,
\begin{equation*}
    f R_i g \iff \int_\Omega u_i ( f (\omega)) dp_i (\omega) \geq \int_\Omega u_i (g (\omega)) dp_i (\omega) . 
\end{equation*}
We call $u_i$  individual $i$'s \textit{value function} and $p_i$ individual $i$'s \textit{belief}.%
\footnote{
Note that while Savage derived finitely additive, convex-ranged measures as individual beliefs, we assume countable additivity and nonatomicity. 
We impose these assumptions in order to apply Lyapunov's convexity theorem (cf.\ Theorem~13.33 of \citet{AB2006Math}) later. 
However, they can be replaced with finite additivity and convex-rangedness if we instead use Theorem~2.2 of \citet{armstrong1981}, a variant of Lyapunov's convexity theorem for finitely additive measures.
}
The set of beliefs is denoted by $\mathcal{P}$. 
For $X\in \mathcal{X}$, the \textit{SEU function} of $R_i$ with $(u_i, p_i)$ is the function $U (\cdot ; u_i, p_i): F_X \rightarrow \mathbb{R}$ defined by, for all $f\in F_X$,
\begin{equation*}
    U (f ; u_i, p_i) = \int_\Omega u_i (f (\omega)) dp_i (\omega).
\end{equation*}
Given $X \in \mathcal{X}$, $\mathcal{R}^\text{SEU} (X)$ denotes the collection of SEU preferences over $F_X$. 
A preference profile $ (R_1, R_2, \ldots, R_n) \in \mathcal{R}^\text{SEU} (X)^N$ is denoted by $R_N$. 
Similarly, a typical element of the set $\mathcal{P}^N$ of  belief profiles is denoted by $p_N = (p_1, p_2, \ldots, p_n)$. 
It is well-known that for $p, q \in \mathcal{P}$ and real-valued functions $u$ and $v$ on $X$, if $U(\cdot ; u, p )$ and $U(\cdot ; v, q )$ represent the same SEU preference, then $p = q$ and $u = \alpha v + \beta$ for some $(\alpha, \beta) \in \mathbb{R}_{++}\times \mathbb{R}$.  

Given $X\in \mathcal{X}$ and $R_N\in \mathcal{R}^\text{SEU} (X)^N$, the \textit{0--1 normalized value function} of individual $i\in N$ is the function $u^\ast_i (\cdot ; X, R_N) : X\to \mathbb{R}$ obtained by applying a positive affine transformation to any value function associated with $R_i$ so that $\max_{x\in X} u^\ast_i (x ; X, R_N) = 1$ and $\min_{x\in X} u^\ast_i (x ; X, R_N) = 0$. 
The \textit{0--1 normalized SEU function} of $R_i$ over $F_X$ is the function $U^\ast_i (\cdot;  X, R_N) \coloneqq U(\cdot;  u^\ast_i (\cdot ; X, R_N), p_i)$. 
Given $X$ and $R_N$, we write $u^\ast (\cdot;  X, R_N) \coloneqq \left(u^\ast_1 (\cdot; X, R_N), \ldots, u^\ast_n (\cdot;  X, R_N)\right)$ for the $n$-tuple of 0--1 normalized value functions, and $U^\ast (\cdot;  X, R_N) \coloneqq \left(U^\ast_1 (\cdot; X, R_N), \ldots, U^\ast_n (\cdot; X, R_N)\right)$ for the $n$-tuple of 0--1 normalized SEU functions. 
In addition, for $R_N$, let $p^\ast(R_i)$ be the unique belief associated with individual $i$'s preference $R_i$ in $R_N$, and $p^\ast(R_N) \coloneqq \left(p^\ast(R_1), \ldots, p^\ast(R_n) \right) \in \mathcal{P}^N$.

A (\textit{social choice}) \textit{problem} is a pair $(X, R_N) $ such that $X\in \mathcal{X}$ and $R_N\in \mathcal{R}^\text{SEU} (X)^N$. 
Let $\mathcal{D}$ be the set of problems. 
An  \textit{aggregation rule} is a function $\mathbf{R}: \mathcal{D} \rightarrow \mathcal{R}$ such that for all $(X, R_N) \in \mathcal{D}$, $\mathbf{R}(X, R_N) \in \mathcal{R} (X)$. 
For $f,g\in F_X$, $f \, \mathbf{R}(X, R_N) \, g$ means that $f$ is at least as socially desirable as $g$. 
Note that the social preference generated by $\mathbf{R}$ does not necessarily satisfy the SEU axioms, unlike the individuals' preferences.
The symmetric and asymmetric parts of $\mathbf{R}(X, R_N)$ are denoted by $\mathbf{I}(X, R_N)$ and $\mathbf{P}(X, R_N)$, respectively.
Given $\mathbf{R}$ and $(X, R_N) \in \mathcal{D}$, we say that $\mathbf{R}(X, R_N)$ is \textit{represented by} a function $W : F_X \to \mathbb{R}$ whenever for all $f,g\in F_X$, $W (f) \geq W(g)$ if and only if $f \, \mathbf{R} ( X, R_N ) \, g$.

\section{Relative fair aggregation rules} \label{sec3}

This section introduces the relative fair aggregation rules, which are the main focus of this paper.
Let $ \Delta_N$ be the set of vectors of nonnegative weights over individuals, that is,  $  \Delta_N = \left\{ \mu \in [0,1 ]^N \mid \sum_{i\in N} \mu_i = 1\right\}$. 

\begin{definition}
    An aggregation rule $\mathbf{R}$ is a \textbf{\textit{relative fair aggregation rule}} if there exists a nonempty, closed, and convex set $\mathcal{M} \subset   \Delta_N$ such that for all $(X,R_N) \in \mathcal{D}$ and all  $f, g \in F_X$,
    \begin{equation}
    \label{eq:def_relfair}
        f \, \mathbf{R}(X, R_N) \, g
        \iff 
        \min_{\mu \in \mathcal{M}} \sum_{i\in N} \mu_i U_i^\ast (f; X, R_N) 
        \geq
        \min_{\mu \in \mathcal{M}} \sum_{i\in N} \mu_i U_i^\ast (g; X, R_N) . 
    \end{equation}
\end{definition}

The relative fair aggregation rule written as \eqref{eq:def_relfair} can be interpreted as the evaluation rule of a social planner who has in mind a set $\mathcal{M}$ of multiple weight vectors. 
The social planner evaluates each act as follows:
First, to provide a natural means of interpersonal comparison, the planner normalizes each individual's utility level to the 0--1 interval. 
For each weight vector $\mu\in \mathcal{M}$, the planner then computes the weighted sum of the normalized utility levels. 
Finally, the planner chooses the minimum weighted sum as the evaluation of that act, which is attained by assigning higher weights to relatively disadvantaged individuals. 

If $\mathcal{M}$ is a singleton, then the relative fair aggregation rule clearly reduces to a weighted utilitarian rule with the 0--1 normalization (e.g., \citet{dhillon1999relative,segal2000let,sprumont2019relative}). 
\begin{definition}\label{def:relativeutilitarian}
    An aggregation rule $\mathbf{R}$ is a \textbf{\textit{relative utilitarian  aggregation rule}} if there exists  $\mu \in   \Delta_N$ such that for all $(X,R_N) \in \mathcal{D}$ and all  $f, g \in F_X$, 
    \begin{equation}
    \label{eq:def_relutili}
        f \, \mathbf{R}(X, R_N) \, g
        \iff 
        \sum_{i\in N} \mu_i U_i^\ast (f;  X, R_N) 
        \geq
        \sum_{i\in N} \mu_i U_i^\ast (g;  X, R_N).  
    \end{equation}
\end{definition}
Notice that the weights $\mu$ over the individuals are exogenously fixed in \eqref{eq:def_relutili}. 
The aggregation rules in Definition \ref{def:relativeutilitarian} are special cases of \citeauthor{sprumont2019relative}'s (\citeyear{sprumont2019relative}) class of relative utilitarian rules, where the weights can vary depending on the belief profile.

On the other hand, if $\mathcal{M}$ is the entire set, then the rule reduces to the Rawlsian maximin rule with the 0--1 normalization. 
\begin{definition}\label{def:relativemaximin}
An aggregation rule $\mathbf{R}$ is a \textbf{\textit{relative maximin aggregation rule}} if, for all $(X,R_N) \in \mathcal{D}$ and all $f, g \in F_X$, 
    \begin{equation}
        f \, \mathbf{R}(X, R_N) \, g
        \iff 
        \min_{i \in N} U_i^\ast (f;  X, R_N) 
        \geq
        \min_{i \in N}  U_i^\ast (g; X, R_N) . 
    \end{equation}
\end{definition}

Hence, the relative fair aggregation rules reconcile the two extreme classes of criteria, the utilitarian rules and the maximin rule, with the 0--1 normalization. 

Finally, we briefly discuss the role of the 0--1 normalization. A primary advantage of adopting a common normalization across problems is that it helps address the indeterminacy of weights in \citeauthor{harsanyi1955}'s (\citeyear{harsanyi1955}) aggregation theorem. As pointed out in the introduction, within Harsanyi's single-profile (or single-problem) framework, individual utility representations can be chosen freely for each problem, which undermines the normative interpretation of the weights assigned to individuals.
In contrast, requiring coherence across multiple problems, as we do in this paper, restricts this arbitrariness. In particular, fixing the range of each agent's utility across problems provides a basis for interpreting the weights in a consistent manner.

The specific range chosen for normalization is not essential: The 0--1 interval is merely one possible choice and is adopted mainly for clarity of exposition.%
\footnote{Using a common range across individuals ensures that each agent's range of experienced satisfaction is measured on a common scale; this common range clarifies the interpretation of our aggregation rules.}
Instead, what matters is that the normalization fixes the range of utility functions rather than relying on alternative approaches, such as scaling by variance or using a poverty line (e.g., \citet{fleurbaey2021fair}).
As we argue in the next section, the range normalization of value functions emerges as a natural choice when imposing reasonable axioms on aggregation rules.

\section{Main characterization} \label{sec4}

This section introduces several axioms and examines their implications step by step to provide an axiomatic foundation for the relative fair aggregation rules.

\subsection{Basic axioms}\label{sec:4.1}

Our basic axioms are about efficiency and continuity. 
The first axiom requires the social planner to respect unanimity among individuals for each problem.%
\footnote{See Section~\ref{subsec:disc_Pareto} for a discussion on this axiom.}

\vspace{1mm}

\begin{description}
    \item[\bf Weak Pareto Principle.] For all $(X, R_N ) \in \mathcal{D}$ and all $f, g\in F_X$, (i) if $f R_i g$ for all  $i\in N$, then $f \, \mathbf{R}(X, R_N) \, g$; and (ii) if $f P_i g$ for all  $i\in N$, then $f \, \mathbf{P}(X, R_N) \, g$. 
\end{description}

\vspace{1mm}

The second axiom postulates Savage's P6 for each problem. 
That is, it requires that the set of states be sufficiently rich to be divided into arbitrarily small events. 

\vspace{1mm}

\begin{description}
    \item[\bf Event Continuity.] For all $(X, R_N) \in \mathcal{D}$, all $f,g\in F_X$ with $f \,\mathbf{P}(X, R_N ) \, g$, and all $x\in X$, there exists a partition $\{ E_k \}_{k=1}^m$ such that $f \,\mathbf{P}(X, R_N ) \, xE_k g$ and $xE_k f \, \mathbf{P}(X, R_N ) \, g$ for all $k \in \{1,2,\ldots, m\}$. 
\end{description}

\subsection{Inter-problem axioms and 0--1 normalization}\label{sec:4-2}

We introduce two inter-problem axioms to ensure the consistency of the social preference across problems.
The first axiom states that individual beliefs play no role in the social evaluation of constant acts because information about individual beliefs is irrelevant when ranking constant acts.

\vspace{1mm}

\begin{description}
    \item[\bf Belief Irrelevance.] For all $(X, R_N), (X, R'_N) \in \mathcal{D}$ such that $u^\ast (\cdot; X, R_N) = u^\ast (\cdot; X, R'_N)$, and all $x, y \in X$, $x \, \mathbf{R}(X, R_N ) \, y$ if and only if $x \, \mathbf{R}(X, R'_N ) \, y$. 
\end{description}

\vspace{1mm}

Note that although the equation $u^\ast (\cdot; X, R_N) = u^\ast (\cdot; X, R'_N)$ is expressed using 0--1 normalized value functions, this axiom does not claim that the social evaluation should be based on the 0--1 normalization of individual utilities. 
Rather, this equation simply means that each individual's taste over outcomes remains unchanged across the two problems $(X, R_N)$ and $(X, R'_N)$.

The next axiom pertains to an independence property with respect to expansions of the feasible set. 
The best-known independence axiom (in the social choice context) is Arrow's independence of irrelevant alternatives. 
An analogous requirement in our framework is formalized as follows: 
For all $(X, R_N ) , (X', R'_N ) \in \mathcal{D}$, if $X\subset X'$ and $R'_N$ coincides with $R_N$ on $F_X$, then $\mathbf{R} (X' , R'_N) $ coincides with $\mathbf{R} (X, R_N )$ on $F_X$.
That is, this property requires that the social evaluation of the original alternatives remain invariant under expansions of the feasible set as long as individual preferences over the originally feasible acts remain unchanged.
Under this axiom, information about how individuals evaluate the outcomes in $X'\backslash X$ is irrelevant for the evaluation of acts whose consequences lie only in $X$. 
As is known in the literature, such strong independence axioms are too restrictive in general:
For instance, \citeauthor{arrow2012social}'s \citeyearpar{arrow2012social} theorem shows the difficulty of imposing the original independence axiom together with the Paretian requirement within a framework where the set of feasible outcomes is fixed.

Rather than imposing independence for every expansion, we restrict its scope to expansions that are ``redundant.''
Informally, an expansion is redundant if, for each new outcome, there is an originally feasible act that gives each individual the same expected utility as that outcome.
The following two examples illustrate why a non-redundant expansion may justify a change in social rankings, whereas a redundant expansion should leave the ranking of originally feasible acts unchanged.

Consider a two-person society with feasible outcomes $x$ and $y$. Suppose that one individual prefers $x$ to $y$, whereas the other prefers $y$ to $x$. 
For the constant acts $x$ and $y$, it would be natural for society to evaluate them as indifferent because the two individuals hold opposite opinions.
Now, suppose that a new outcome $z$ becomes available, and the first individual strictly prefers it the most, whereas the second strictly prefers it the least.
Clearly, Arrow's independence axiom requires that society remain indifferent between $x$ and $y$ after the expansion. 
However, in the new problem, $x$ is a middle-of-the-road alternative for both individuals in the sense that they rank it second.
Thus, the intermediate position of $x$ could justify society ranking it above $y$.
This suggests that the expansion is not an irrelevant change for society---$z$ gives utility levels that were not achievable in the original problem, and adding $z$ may significantly change the resulting relative positions of $x$ and $y$. 

By contrast, suppose a new outcome $z'$, which gives each individual the average utility level of $x$ and $y$, becomes feasible in place of $z$. 
Then, since there is an event $E$ that both individuals believe will occur with probability $1/2$ (according to Lyapunov's convexity theorem), the originally feasible act $xEy$ gives both individuals the same expected utility level of $z'$.
In this case, the new outcome $z'$ is redundant, and hence, there is no reason for society to change the preference over the acts constructed from the originally feasible outcomes (i.e., $x$ and $y$).

We now formalize the type of expansions to which our independence axiom applies.
For two problems $(X,R_N), (X', R'_N) \in\mathcal{D}$, we say that 
$(X', R'_N)$ is a \textit{redundant-outcome expansion of} $(X, R_N)$ 
if (i) $X\subset X'$, (ii) $R'_N$ coincides with $R_N$ on $F_X$, and (iii) for all $x'\in X'$, there exists $f\in F_X$ such that $fI'_i x'$ for all $i\in N$. 
The following axiom requires that under redundant-outcome expansions, the social evaluation of the original acts remain unchanged.

\vspace{1mm}

\begin{description}
    \item[\bf Independence of Redundant Outcomes.] For all $(X, R_N) , (X', R'_N ) \in \mathcal{D}$ such that $(X', R'_N)$ is a redundant-outcome expansion of $(X, R_N)$, $\mathbf{R} (X' , R'_N) $ coincides with $\mathbf{R} (X, R_N )$ on $F_X$.
\end{description}

\vspace{1mm}

\citet{dhillon1999relative} studied the corresponding axiom in the context of collective decisions under risk. 
Furthermore, in the present framework, \citet{sprumont2019relative} discussed the above axiom to explain their independence axiom (see the discussion after Lemma~\ref{lem:basic}).
\citet{brandl2021belief} also examined a similar axiom in a different framework for preference aggregation under uncertainty.

The first result states that an aggregation rule satisfies the four preceding axioms if and only if the social planner evaluates acts by ranking their 0--1 normalized utility vectors with a ``social welfare'' function. 
We say that a function $\psi: [0,1]^N \rightarrow \mathbb{R}$ is \textit{monotonic} if $\psi (\mathbf{u}) > \psi (\mathbf{v} )$ for all $\mathbf{u}, \mathbf{v} \in [0,1]^N$ such that $\mathbf{u} \gg \mathbf{v}$.\footnote{For any vectors $\mathbf{u}=(\mathbf{u}_1,\ldots,\mathbf{u}_n)\in\mathbb{R}^N$ and $\mathbf{v}=(\mathbf{v}_1,\ldots,\mathbf{v}_n)\in\mathbb{R}^N$, we write $\mathbf{u} \gg \mathbf{v}$ if $\mathbf{u}_i > \mathbf{v}_i$ for all $i \in N$, and $\mathbf{u} \geq \mathbf{v}$ if $\mathbf{u}_i \geq \mathbf{v}_i$ for all $i \in N$. } 

\begin{lem}\label{lem:basic}
    An aggregation rule $\mathbf{R}$ satisfies \textnormal{weak Pareto principle}, \textnormal{event continuity}, \textnormal{belief irrelevance}, and \textnormal{independence of redundant outcomes} 
    if and only if there exists a monotonic continuous function $\psi :[0,1]^N \rightarrow \mathbb{R}$ such that
    for each $(X, R_N) \in \mathcal{D}$, $\mathbf{R} (X, R_N)$ is represented by the function $W_{(X, R_N)} :F_X \to \mathbb{R}$ defined by, for all $f\in F_X$,
    \begin{equation*}
        W_{(X, R_N)} (f) = \psi \big(U^\ast (f;  X, R_N)\big). 
    \end{equation*}  
\end{lem}

This result provides guidance regarding what information the social evaluation should rely on: 
Under the axioms in the above lemma, only the 0--1 normalized utility levels matter. 

Note that \citet{sprumont2019relative} also derived the 0--1 normalization using a version of the Pareto principle and an independence axiom stronger than \textit{independence of redundant outcomes}. 
It requires that the social preference remain invariant under the expansion of the feasible set if, for each individual, the new outcomes do not change the best and worst outcomes in the original feasible set.%
\footnote{
\label{fn:IIE}
Formally, this axiom, \textit{independence of inessential expansion}, is defined as follows: For all $(X, R_N ) , (X', R'_N ) \in \mathcal{D}$, if (i) $X\subset X'$, (ii) $R'_N$ coincides with $R_N$ on $F_X$, and (iii) for all $x' \in X'$ and all $i\in N$, there exist $x^+_i, x^-_i \in X$ such that $x^+_i R'_i x' R'_i x^-_i$, then $\mathbf{R} (X' , R'_N) $ coincides with $\mathbf{R} (X, R_N )$ on $F_X$.
}
Since the result does not rely on \textit{belief irrelevance}, the derived ranking over normalized vectors can vary depending on the belief profile.
In contrast, our first lemma shows that, under \textit{belief irrelevance}, the 0--1 normalization is obtained from a weaker independence axiom.

\subsection{Preference for mixing and inequality aversion}
\label{subsec:4-3}

In this part, we introduce the intra-problem axiom under which tossing a fair coin to choose between two outcomes should be weakly more desirable for society than committing to one of the deterministic outcomes. 

This axiom is central to addressing the second issue concerning \citeauthor{harsanyi1955}'s (\citeyear{harsanyi1955}) aggregation theorem discussed in Section~\ref{sec1}. 
\cite{diamond} pointed out that Harsanyi's utilitarian aggregation rules are insensitive to the distribution of opportunities to be better off across individuals. 
To illustrate this critique in our framework, consider a society consisting of two individuals, $i$ and $j$, 
who are informed that an event $E$ occurs with probability $1/2$. 
Consider three ``risky'' acts that yield outcomes depending on whether the realized state belongs to $E$ or not. 
Table~\ref{tab:diamond} displays each individual's normalized utility levels under these acts. 
The constant act $f^1$ gives individual $i$ utility $1$ and individual $j$ utility $0$ with certainty, whereas the constant act $f^2$ reverses these utilities.
In contrast, the act $g$ yields the outcomes of $f^1$ and $f^2$ with equal probability.
Thus, $g$ gives each individual an equal chance of obtaining the higher utility level.
From the perspective of ex-ante fairness, $g$ would be more socially desirable than $f^1$ and $f^2$. 
However, an equal-weighted utilitarian rule, which is a typical Harsanyian aggregation rule, is indifferent between them because each act generates the same weighted sum of utility levels. 
More generally, no utilitarian rule can always prioritize an act with equal opportunities in such situations.

\begin{table}[t]
    \caption{Three acts: with ex-ante fairness (act $g$) and without it (acts $f^1$ and $f^2$)} 
    \label{tab:diamond}
    \centering
    \begin{subtable}{.3\linewidth}
      \centering
        \caption{}\label{tab:unfair}
        \begin{tabular}{rcc}
            \toprule
            $f^1$ & $E$ & $E^c$ \\
             \midrule 
            $i$ & $1$ & $1$ \\
            $j$ & $0$ & $0$ \\
             \bottomrule
        \end{tabular}
    \end{subtable} 
    \begin{subtable}{.3\linewidth}
      \centering
        \caption{}\label{tab:unfair2}
        \begin{tabular}{rcc}
            \toprule
            $f^2$ & $E$ & $E^c$ \\
             \midrule 
            $i$ & $0$ & $0$ \\
            $j$ & $1$ & $1$ \\
             \bottomrule
        \end{tabular}
    \end{subtable}
    \begin{subtable}{.3\linewidth}
      \centering
        \caption{}\label{tab:fair}
        \begin{tabular}{rcc}
            \toprule
            $g$ & $E$ & $E^c$ \\
             \midrule 
            $i$ & $1$ & $0$ \\
            $j$ & $0$ & $1$ \\
            \bottomrule
        \end{tabular}
    \end{subtable}%
\end{table}

Motivated by the concern about utilitarianism under risk, we formalize the essence of \citeauthor{diamond}'s (\citeyear{diamond}) point as an axiom in our framework. 
Take $(X, R_N) \in \mathcal{D}$ arbitrarily, and let $p_N$  denote $p^\ast (R_N)$. We say that an event $E\subset \Omega$ is a \textit{coin-toss event} (in the problem $(X, R_N)$) if  $p_i (E) = 1/2$ for all $i\in N$. 
Note that by applying Lyapunov's convexity theorem, 
the existence of coin-toss events can be ensured in all problems. 
Our first intra-problem axiom is formalized as follows:

\vspace{1mm}

\begin{description}
    \item[\bf Weak Preference for Mixing.] For all $(X, R_N) \in \mathcal{D}$, all $x,y\in X$, and all coin-toss events $E\subset \Omega$, $xEy\,  \mathbf{R} (X, R_N ) \, x$ or $xEy \,  \mathbf{R} (X, R_N ) \, y$. 
\end{description}

\vspace{1mm}

The next lemma shows that \textit{weak preference for mixing}, together with the axioms in Lemma~\ref{lem:basic}, implies that the function $\psi$ in Lemma~\ref{lem:basic} satisfies quasiconcavity, which captures the fairness concern of the social planner.

\begin{lem}\label{lem:quasiconcav}
    An aggregation rule $\mathbf{R}$ satisfies \textnormal{weak Pareto principle}, \textnormal{event continuity}, \textnormal{belief irrelevance},  \textnormal{independence of redundant outcomes},  and \textnormal{weak preference for mixing}  
    if and only if there exists a monotonic, continuous, quasiconcave function $\psi :[0,1]^N \rightarrow \mathbb{R}$ such that
    for each $(X, R_N) \in \mathcal{D}$, $\mathbf{R} (X, R_N)$ is represented by the function $W_{(X, R_N)} :F_X \to \mathbb{R}$ defined by $W_{(X, R_N)} (f) = \psi \big(U^\ast (f;  X, R_N)\big)$ for all $f\in F_X$.
\end{lem}

To understand the role of \textit{weak preference for mixing} in detail, we compare this lemma with \citeauthor{sprumont2019relative}'s (\citeyear{sprumont2019relative}) characterization of the (belief-weighted) relative utilitarian aggregation rules.
In \citet{sprumont2019relative}, the central axiom in the derivation of additive representation is Savage's P2, an axiom of eventwise separability.\footnote{
Savage's P2, or \textit{sure-thing principle} in \cite{sprumont2019relative}, requires that for all $(X, \mathcal{R}_N) \in \mathcal{D}$, all $E\subset \Omega$, and all $f,f', g, g' \in F_X$ such that (i) $f(\omega) = f' (\omega)$ and  $g(\omega) = g' (\omega)$ for all $\omega\in E$; and (ii)  $f(\omega) = g (\omega)$ and $ f'(\omega) = g' (\omega)$ for all $\omega\in \Omega \backslash E$, we have $f \,  \mathbf{R} (X, R_N ) \, g $ if and only if $f' \,  \mathbf{R} (X, R_N ) \, g'$. 
}
This axiom effectively renders the social preference insensitive to ex-ante fairness.
To see this, consider an aggregation rule $\mathbf{R}$ that satisfies Savage's P2 and the axioms in Lemma~\ref{lem:basic}. 
Given $(X,R_N)\in\mathcal{D}$, take $x,y \in X$ such that $x\,\mathbf{I}(X,R_N)\,y$, and consider $xEy \in F_X$, where $E$ is a coin-toss event. 
Assume, for expositional purposes, that $xEy \,\mathbf{P}(X, R_N)\, y$ holds.
Because Savage's P2 suggests that $xEy \,\mathbf{P}(X, R_N)\, y$ if and only if $x \,\mathbf{P}(X, R_N)\, yEx$,
we have $xEy\,\mathbf{P}(X,R_N)\,yEx$.
However, since $E$ is a coin-toss event, all individuals are indifferent between $xEy$ and $yEx$. 
Then, \textit{weak Pareto principle} implies $xEy\,\mathbf{I}(X,R_N)\,yEx$, 
which is a contradiction. 

This argument demonstrates that aggregation rules cannot strictly prefer mixing to not mixing while adhering to Savage's P2.
To avoid being constrained to the class of utilitarian rules, we adopt \textit{weak preference for mixing} instead of Savage's P2. 

\subsection{Certainty independence without objective randomization} \label{subsec:4-4}

We consider the second intra-problem axiom, which concerns mixture independence.
In decision theory, the independence condition in the vNM theorem and its variants have been studied as one of the individual rationality assumptions. 
Here, we focus on a weak version of the independence axiom under uncertainty, known as ``certainty independence'' (\citet{gilboa1989maxmin}), as a condition for social preferences.
This axiom requires that the ranking of any two acts should be invariant when a common constant act is mixed with each original act in the same proportion.
Since mixing constant acts decreases the degree of uncertainty of the two original alternatives in the same proportion, this affects both acts in a similar way. 

These mixture operations are allowed \textit{a priori} in \citet{gilboa1989maxmin} because their analysis is conducted in the Anscombe--Aumann setting, where the existence of a randomization device is assumed by modeling outcomes as lotteries. 
In contrast, such mixture operations are not allowed in our framework. 
To address this limitation, we utilize individuals' beliefs to define acts that can be regarded as if they were generated by a randomization device.
For $p_N \in \mathcal{P}^N$ and $E \subset \Omega$, let $p_N(E) = (p_1 (E), p_2 (E), \ldots, p_n(E))$. 
For any $(X, R_N) \in \mathcal{D}$ with $p_N = p^\ast (R_N)$, any $f, g\in F_X$, and any $\alpha \in (0,1)$, we define a \textbf{\textit{pseudo-mixed act}} of $f$ and $g$, denoted by $f_{\alpha} g$, by the condition that, for all $x \in X$, 
\begin{equation}
\label{eq:def_ps_mix}
    p_N \left((f_{\alpha}g)^{-1} (x)\right) = 
    \alpha \cdot p_N \left(f^{-1} (x)\right) + \left(1- \alpha\right)\cdot p_N \left(g^{-1} (x)\right) . 
\end{equation} 
That is, $f_{\alpha} g$ is an act that all individuals believe assigns probability $\alpha$ to $f$ and probability $1-\alpha$ to $g$. 

One might doubt that acts that satisfy the condition in \eqref{eq:def_ps_mix} always exist.
However, using Lyapunov's convexity theorem, we can ensure the existence of pseudo-mixed acts in our model.%
\footnote{We thank an anonymous referee for suggesting the current statement of Proposition~\ref{prop:pmact} and its proof.} 

\begin{proposition}\label{prop:pmact}
    For all $(X, R_N) \in \mathcal{D}$, all  $f,g \in F_X$, and all $\alpha\in (0,1)$, there exists a pseudo-mixed act $f_{\alpha} g$. 
\end{proposition}

Note that although $f_{\alpha} g$ is not uniquely determined in general, the argument in this paper does not depend on any specific choice of pseudo-mixed acts.

Using the above definition, we introduce our weak version of the independence axiom.

\vspace{1mm}

\begin{description}
    \item[\bf Restricted Certainty Independence.] For all $(X, R_N) \in \mathcal{D}$ such that $u_i^\ast (\cdot; X, R_N) = u_j^\ast (\cdot; X, R_N)$ for each $i,j\in N$, all $f, g \in F_X$, all $x\in X$, and all $\alpha \in (0,1)$, $f \, \mathbf{R}(X, R_N ) \,   g$ if and only if $f_{\alpha} x \, \mathbf{R}(X, R_N ) \, g_{\alpha} x$. 
\end{description}

\vspace{1mm}

Notice that \textit{restricted certainty independence} focuses on cases where all individuals share a common value function (i.e., they have the same taste over outcomes).
To see the role of this restriction, consider two acts $f$ and $g$ in a problem with two individuals who have different value functions.  
Suppose that the 0--1 normalized utility vectors $(1, 0)$ and $(0, 1)$ are realized in $f$ and $g$, respectively, and that these two acts are equally desirable for society. 
Take an outcome $x$ that yields the normalized utility vector $(1, 0)$ and compare the two mixed acts $f_{\alpha} x$ and $g_{\alpha} x$ with $\alpha = 1/2$.
If we impose the certainty independence condition without the restriction in this case, then $f_{\alpha} x$ and $g_{\alpha} x$ should remain equally desirable. 
However, since $f_{\alpha} x$ yields $(1, 0)$ and $g_{\alpha} x$ yields $(1/2, 1/2)$, the mixture operation heterogeneously affects the utility distributions. 
In such a case, the ranking of the original acts does not need to be preserved, and as a result, the independence property is less compelling.

In contrast, when all individuals share a common value function, each constant act yields a constant normalized utility vector.
Hence, \textit{restricted certainty independence} applies only when the pseudo-mixture operation affects all individuals uniformly. 
Note that the condition that the act mixed with the original acts be constant plays a similar role:
Without this restriction, the mixture operation affects individuals heterogeneously due to differences in beliefs.

The following lemma characterizes the implications of \textit{restricted certainty independence} when combined with the axioms in Lemma~\ref{lem:basic}.
We say that a function $\psi: [0,1]^N \rightarrow \mathbb{R}$ is \textit{homogeneous} if $\psi (\alpha \mathbf{u})  = \alpha \psi (\mathbf{u} )$ for all $\mathbf{u} \in [0,1]^N$ and  all $\alpha > 0$ such that $\alpha \mathbf{u}  \in [0,1]^N$; 
a function is \textit{translation-invariant} if $\psi (\mathbf{u} + c \mathbf{1})  = \psi (\mathbf{u} ) + c$ for all $\mathbf{u} \in [0,1]^N$ and all $c \in \mathbb{R}$ such that $ \mathbf{u} + c \mathbf{1}  \in [0,1]^N$.%
\footnote{Let $\mathbf{0}$ and $\mathbf{1}$ denote the constant vectors $(0,0,\ldots, 0) $ and $(1,1,\ldots, 1)$ in $[0,1]^N$, respectively.}

\begin{lem}
\label{lem:4addCI}
    An aggregation rule $\mathbf{R}$ satisfies \textnormal{weak Pareto principle}, \textnormal{event continuity}, \textnormal{belief irrelevance}, \textnormal{independence of redundant outcomes}, and \textnormal{restricted certainty independence}
    if and only if 
    there exists a monotonic, continuous, homogeneous, and translation-invariant function $\psi :[0,1]^N \rightarrow \mathbb{R}$ such that
    for each $(X, R_N) \in \mathcal{D}$, $\mathbf{R} (X, R_N)$ is represented by the function $W_{(X, R_N)} :F_X \to \mathbb{R}$ defined by $W_{(X, R_N)} (f) = \psi \big(U^\ast (f;  X, R_N)\big)$ for all $f\in F_X$.
\end{lem}

It is worth noting that \textit{belief irrelevance} plays a crucial role in this result.
To see this, let $\succsim$ be the binary relation on $[0,1]^N$ represented by the function $\psi$ in Lemma~\ref{lem:basic}. 
Take $c\in [0,1]$, $\mathbf{u}, \mathbf{v}\in [0, 1]^N$, and $\alpha \in (0,1)$ arbitrarily.
In the proof of Lemma~\ref{lem:4addCI}, we prove that $\mathbf{u}\succsim \mathbf{v} $ if and only if $\alpha\mathbf{u} + (1- \alpha) c\mathbf{1} \succsim \alpha \mathbf{v} + (1- \alpha) c\mathbf{1}$, using \textit{restricted certainty independence}. 
In this process, we construct a problem $(X, R_N) \in \mathcal{D}$ in which all individuals share a common value function, and there exists $(f,g,x) \in F_X \times F_X \times X$ whose normalized utility vectors correspond to $\mathbf{u}$, $\mathbf{v}$, and $c\mathbf{1}$, respectively. 
This construction is possible because Lemma~\ref{lem:basic}, which involves \textit{belief irrelevance}, allows us to choose any belief profile.
If we instead attempted to obtain the properties in Lemma~\ref{lem:4addCI} under each belief profile, then we would need to construct such a problem and such acts while keeping the belief profile fixed.
However, this is not possible for every belief profile under the restriction that all individuals share a common value function.
For instance, if all individuals' beliefs coincide, then the normalized utility vectors associated with all acts lie only on the diagonal line.

\subsection{Characterization of relative fair aggregation rules}\label{sec:4-5} 

We now state our main result. The relative fair aggregation rules can be characterized by the axioms employed in the lemmas above.

\begin{theorem}
\label{thm_relativefair}
    An aggregation rule $\mathbf{R}$ satisfies \textnormal{weak Pareto principle}, \textnormal{event continuity}, \textnormal{belief irrelevance}, \textnormal{independence of redundant outcomes}, \textnormal{weak preference for mixing}, and \textnormal{restricted certainty independence}
    if and only if it is a relative fair aggregation rule. 
\end{theorem}

As shown in Lemma~\ref{lem:basic}, the two inter-problem axioms, \textit{belief irrelevance} and \textit{independence of redundant outcomes}, yield the 0--1 normalization. 
The structure of the evaluation function defined over the 0--1 normalized utility vectors is mainly obtained from \textit{restricted certainty independence} and \textit{weak preference for mixing} 
(note, once again, that \textit{belief irrelevance} plays an important role, as discussed in Section~\ref{subsec:4-4}). 
By combining these intra-problem axioms with the inter-problem ones, the above theorem characterizes the relative fair aggregation rules, which can address the issues of \citeauthor{harsanyi1955}'s (\citeyear{harsanyi1955}) theorem discussed in Section~\ref{sec1}. 

A primary theoretical contribution of our theorem is to provide, within the Savage setting, a characterization of aggregation rules that have a functional form akin to \citeauthor{gilboa1989maxmin}'s (\citeyear{gilboa1989maxmin}) maxmin EU model. 
While the maxmin EU model has been axiomatized within the Savage framework mainly in the context of individual decision-making, the axioms involved are somewhat complex (e.g., \citet{casadesus2000maxmin,ghirardato2003subjective,alon2014purely,borie2023maxmin}).
In contrast to these results, our theorem derives a similar functional form from relatively simple axioms, some of which rely on the pseudo-mixture operation.
This operation is possible precisely because we consider the aggregation problem, in which we can utilize information about individual preferences.
Note that even in the Anscombe--Aumann setting, the corresponding characterization of the relative fair aggregation rules can be established by applying a similar argument.

In Appendix~\ref{app_ind}, we verify the independence of the axioms in Theorem~\ref{thm_relativefair}.
Furthermore, in Appendix~\ref{app_general}, we study a generalization of relative fair aggregation rules. 
The characterization of these rules also relies on the mixture operation proposed in this paper. 

In the remainder of this section, we discuss the relation between the impartiality axiom and the properties of the associated weight set. 
We call a bijective function $\pi: N\to N$ a \textit{permutation}. 
The set of permutations is denoted by $\Pi$. 
For $(X, R_N) \in \mathcal{D}$ and $\pi\in \Pi$, let $R_N^\pi$ be the profile $(R^\pi_1, \ldots, R^\pi_n)$ such that for all $i\in N$, $R^\pi_i = R_{\pi (i)}$. 
The axiom of impartiality is formalized as follows:

\vspace{1mm}

\begin{description}
\item[Anonymity.] For all $(X,R_N) \in \mathcal{D}$ and all $\pi \in \Pi$, $\mathbf{R}(X,R_N) = \mathbf{R}(X, R_N^\pi)$.
\end{description}

\vspace{1mm}

As usual, the symmetry of social welfare functions can be derived from this axiom. 
We say that a function $\psi: [0,1]^N \to \mathbb{R}$ is \textit{symmetric} if, for all $\mathbf{u} \in [0,1]^N$ and all $\pi \in \Pi$,  $\psi(\mathbf{u}) = \psi(\mathbf{u}^\pi)$, where $\mathbf{u}^\pi=(\mathbf{u}_{\pi(1)},\ldots,\mathbf{u}_{\pi(n)})$.

\begin{lem}\label{lem:addAnonym}
    An aggregation rule $\mathbf{R}$ satisfies \textnormal{weak Pareto principle}, \textnormal{event continuity}, \textnormal{belief irrelevance}, \textnormal{independence of redundant outcomes},  and \textnormal{anonymity} 
    if and only if 
    there exists a symmetric, monotonic, and continuous function $\psi :[0,1]^N \rightarrow \mathbb{R}$ such that
    for each $(X, R_N) \in \mathcal{D}$, $\mathbf{R} (X, R_N)$ is represented by the function $W_{(X, R_N)} :F_X \to \mathbb{R}$ defined by $W_{(X, R_N)} (f) = \psi \big(U^\ast (f;  X, R_N)\big)$ for all $f\in F_X$.
\end{lem}

We say that a set $\mathcal{M} \subset \Delta_N$ is \textit{symmetric} if $\mu^\pi \in \mathcal{M}$ for all $\mu \in \mathcal{M}$ and all $\pi\in\Pi$. 
By combining this lemma with Theorem~\ref{thm_relativefair}, we can easily characterize the class of relative fair aggregation rules with a symmetric weight set. 

\begin{corollary}
\label{cor:sym}
An aggregation rule $\mathbf{R}$ satisfies \textnormal{weak Pareto principle}, \textnormal{event continuity}, \textnormal{belief irrelevance}, \textnormal{independence of redundant outcomes}, \textnormal{weak preference for mixing},  \textnormal{restricted certainty independence}, and \textnormal{anonymity}
if and only if it is a relative fair aggregation rule associated with a symmetric set $\mathcal{M}$. 
\end{corollary}

\section{Special cases} \label{sec5}
We examine two special cases of the relative fair aggregation rules: the relative utilitarian and the relative maximin aggregation rules.
The relative utilitarian aggregation rules are characterized by the axioms in Lemma~\ref{lem:basic} and an independence axiom stronger than \textit{restricted certainty independence}.
On the other hand, the relative maximin aggregation rule is obtained from the axioms in Lemma~\ref{lem:basic}, \textit{anonymity}, and a stronger version of \textit{weak preference for mixing}.
Therefore, roughly speaking, each of the two extreme cases can essentially be derived from the characterization of the relative fair aggregation rules by strengthening one of the intra-problem axioms while dropping the other.
Moreover, we provide an alternative characterization of the relative maximin aggregation rule using an axiom of collective attitudes toward uncertainty. 

\subsection{Relative utilitarian aggregation rules}

First, we examine the relative utilitarian aggregation rules at the axiomatic level.
Recall that \textit{restricted certainty independence} requires that if all individuals share a common 0--1 normalized value function, then pseudo-mixing constant acts in a common proportion should not change social rankings. 
By focusing on the situations where tastes coincide, we can fix the effect of mixing constant acts among individuals (otherwise, mixing some constant act may be preferable for some individuals but not for others). 
This axiom is rooted in \citeauthor{gilboa1989maxmin}'s (\citeyear{gilboa1989maxmin}) axiom of certainty independence, but the original was defined without any restriction on tastes since they studied individual decision-making.
In our setup, the direct counterpart of their axiom is defined as follows: 

\vspace{1mm}

\begin{description}
    \item[\bf Certainty Independence.] For all $(X, R_N) \in \mathcal{D}$, all $f, g \in F_X$, all $x\in X$, and all $\alpha \in (0,1)$,
    $f \, \mathbf{R}(X, R_N ) \,  g$ if and only if $f_{\alpha} x \, \mathbf{R}(X, R_N ) \, g_{\alpha} x$. 
\end{description}

\vspace{1mm}

The next theorem shows that if we replace \textit{restricted certainty independence} in Theorem~\ref{thm_relativefair} with \textit{certainty independence}, then we obtain the relative utilitarian aggregation rules. 
Here, \textit{weak preference for mixing} can be dropped since it becomes redundant. 

\begin{theorem}
\label{thm:rel-util}
    An aggregation rule $\mathbf{R}$ satisfies \textnormal{weak Pareto principle}, \textnormal{event continuity}, \textnormal{belief irrelevance}, \textnormal{independence of redundant outcomes}, and \textnormal{certainty independence}
    if and only if 
    it is a relative utilitarian aggregation rule. 
\end{theorem}

Note that full independence is not necessary to obtain relative utilitarian aggregation rules. 
Intuitively, full independence requires that, when comparing two alternatives, mixing each of them with a third alternative in the same proportion should not change their relative desirability. 
This property has played a central role in characterizing additive representations in the literature (e.g., \citet{anscombe1963definition,dAspremont1977equity}). 
In comparison to it, \textit{certainty independence}, which is defined by utilizing the framework with uncertainty, is weaker in the sense that it imposes the same requirement only when the third alternative is a constant act.

In the following, we explain why relative utilitarian aggregation rules can be obtained even with the weaker independence condition. 
Let $\mathbf{R}$ be an aggregation rule that satisfies the axioms in Theorem~\ref{thm:rel-util}.
By Lemma~\ref{lem:basic}, there is a function $\psi:[0,1]^N \to \mathbb{R}$ induced by $\mathbf{R}$.
In the normalized vector space $[0,1]^N$, additive representations are closely connected to the following independence property: For all $\mathbf{u}, \mathbf{v}, \mathbf{w} \in [0,1]^N$ and all $\alpha\in (0,1)$, $\psi (\mathbf{u}) \geq \psi (\mathbf{v})$ if and only if $\psi (\alpha \mathbf{u} + (1-\alpha )\mathbf{w}) \geq \psi (\alpha \mathbf{v} + (1-\alpha )\mathbf{w})$. 
To show that this property is satisfied, we take $(X, R_N) \in\mathcal{D}$ and $(f,g,x) \in F_X \times F_X \times X$ such that their normalized utility vectors correspond to $\mathbf{u}$, $\mathbf{v}$, and $\mathbf{w}$, respectively (which is impossible as long as we restrict attention to problems with taste agreement, as in \textit{restricted certainty independence}).
By applying \textit{certainty independence} to $(f,g,x)$, we obtain the above property over the normalized vector space. 
Therefore, \textit{certainty independence} yields an additive representation of the social welfare function on the normalized vector space. 

\subsection{Relative maximin aggregation rule}

The relative maximin aggregation rule is the other extreme case of the relative fair aggregation rules. 
Here, we provide axiomatic foundations for this case. 

In the characterization of the relative fair aggregation rules, \textit{weak preference for mixing} captures the egalitarian attitude.
This axiom requires that for any two outcomes $x$ and $y$, an uncertain act in which each of them is realized with probability $1/2$ be weakly better than at least one of the original outcomes.
Such randomization is implemented by constructing an act $xEy$, where $E$ is a coin-toss event.

We now introduce an axiom that does not impose any restrictions on the events used in mixing outcomes.  
\vspace{1mm}

\begin{description}
    \item[\bf Strong Preference for Mixing.] For all $(X, R_N) \in \mathcal{D}$, all $x,y\in X$, and all $E\subset \Omega$, $xEy\, \mathbf{R}(X, R_N ) \, x$ or $xEy \,  \mathbf{R} (X, R_N) \, y$. 
\end{description}

\vspace{1mm}

Since this axiom requires that \textit{any} compromise between two outcomes be weakly more desirable than at least one of them, 
it implies a stronger egalitarian attitude than \textit{weak preference for mixing} does.
The following result shows that, together with the axioms in Lemma~\ref{lem:basic} and \textit{anonymity}, \textit{strong preference for mixing} implies that the social planner evaluates acts in the fairest way; that is, the relative maximin aggregation rule is derived. 
Note that \textit{anonymity} is indispensable for the following result to rule out aggregation rules such as dictatorship. 

\begin{theorem}
\label{thm_relmaxmin_mix}
    An aggregation rule $\mathbf{R}$ satisfies \textnormal{weak Pareto principle}, \textnormal{event continuity}, \textnormal{belief irrelevance},  \textnormal{independence of redundant outcomes}, \textnormal{anonymity}, and \textnormal{strong preference for mixing}
    if and only if 
    it is the relative maximin aggregation rule. 
\end{theorem}

From the theorems obtained so far, we can see that the key implications of the relative utilitarian and relative maximin aggregation rules can be captured by strengthening one of the intra-problem axioms in Theorem~\ref{thm_relativefair} (and dropping the other). 
This relationship highlights how the intra-problem axioms in Theorem~\ref{thm_relativefair} relate to the key concepts of utilitarianism and egalitarianism and clarifies the role of restriction in these axioms.

In the remainder of this subsection, we provide another axiomatic foundation for the relative maximin aggregation rule using an axiom of uncertainty aversion employed in studies such as \citet{gilboa2010objective} and \citet{alon2016utilitarian}. 
The following axiom requires that for any problem where individuals share a common value function, the social planner avoid uncertain alternatives if some individual does so. 
In other words, this postulates that all individuals have veto power to block ambiguous acts in favor of unambiguous ones.

\vspace{1mm}

\begin{description}
    \item[\bf Social Ambiguity Avoidance.] For all $(X, R_N) \in \mathcal{D}$ such that $u_i^\ast (\cdot; X, R_N) = u_j^\ast (\cdot; X, R_N)$ for each $i, j \in N$, all $f\in F_X$, and all $x\in X$, if there exists $k\in N$ such that $x P_k f$, then $x \, \mathbf{P} (X, R_N) \, f$. 
\end{description}

\vspace{1mm}
Restricting attention to preference profiles in each of which all individuals share a common value function implies that each constant act yields the same normalized expected utility across individuals.
Since \textit{social ambiguity avoidance} prioritizes constant acts, it drives social evaluation toward equality.
Indeed, the theorem below characterizes the relative maximin aggregation rule by combining this axiom with those in Lemma~\ref{lem:basic}.

\begin{theorem}
\label{thm_ambavo_maxmin}
    An aggregation rule $\mathbf{R}$ satisfies \textnormal{weak Pareto principle}, \textnormal{event continuity}, \textnormal{belief irrelevance}, \textnormal{independence of redundant outcomes}, and \textnormal{social ambiguity avoidance}
    if and only if 
    it is the relative maximin aggregation rule. 
\end{theorem}

This result is closely related to Theorem~4 in \citet{gilboa2010objective}. 
They adopted corresponding axioms in the context of individual decision-making under uncertainty and constructed the most cautious decision criterion from an incomplete preference in the agent's mind. 
In contrast, we characterize the egalitarian aggregation rule based on the incomplete ranking induced by \textit{weak Pareto principle}.

\section{Discussion} \label{sec6}

To conclude this paper, we discuss the relationship between our results and those in the literature. 

\subsection{Spurious unanimity and  the Pareto principle}
\label{subsec:disc_Pareto}
 
As mentioned in Section~\ref{sec1}, if both individuals and the social planner have SEU preferences, the full Pareto principle leads to dictatorship under belief disagreement (\citet{hylland1979impossibility,mongin1995consistent,mongin1998paradox}). 
This tension stems from the conflict between the Pareto principle and the statewise monotonicity axiom, the latter being a presumption of the SEU model (\citet{chambershayashi2006preference}). 
Previous works have avoided this impossibility either by weakening the Pareto principle or by dropping statewise monotonicity while retaining the Pareto principle.
This paper adopts the second approach. 

The first approach is based on \citeauthor{mongin1998paradox}'s (\citeyear{mongin1998paradox}) criticism of the Pareto principle under uncertainty. 
Mongin pointed out that the full Pareto principle is not so appealing under uncertainty since agreements may reflect \textit{spurious unanimity}---that is, unanimity resulting from a combination of disagreements about beliefs and tastes.
Motivated by this criticism, several papers have introduced weaker versions of the Pareto principle and examined their implications (e.g., \citet{gilboa2004utilitarian,alon2016utilitarian,danan2016robust}).\footnote{Alternatively, some papers have focused on the case where all individuals have the same tastes over constant outcomes to avoid the problem of respecting spurious unanimity (e.g., \citet{cres2011aggregation,stanca2021smooth}).}

One might think that the relative fair aggregation rules are undesirable since these rules satisfy the full Pareto principle and are thus vulnerable to Mongin's concern. 
However, as \citet{sprumont2018belief} pointed out, dropping the Pareto principle is ``dangerous''; at least, it would be undesirable to reject some rule just because it satisfies the Pareto principle under uncertainty.
Since a subjective probability distribution in Savage's SEU model is ``just an abstract system of weights'' obtained as an implication of a series of axioms, beliefs are also part of individual tastes and therefore should be respected by society.
Furthermore, the former approach has an important limitation.
Aggregation rules that satisfy statewise monotonicity are subject to \citeauthor{diamond}'s (\citeyear{diamond}) criticism of utilitarianism, discussed in Section~\ref{subsec:4-3}, and therefore cannot accommodate considerations of equality of opportunity. 
Exploring aggregation rules that can address the concerns of both approaches is left for future research.%
\footnote{For a few exceptions, see \citet{ben1997measurement}, \citet{gajdosmaurin2004unequal}, and \citet{mongin2021rawls}, where individual preferences are not explicitly modeled, and uncertain alternatives are directly formalized as a function that assigns an income distribution to each state.}

\subsection{Comparison with Sprumont's results} \label{sec6:sprumont}

We discuss the connection between a series of four papers by \citet{sprumont2013relative,sprumont2018belief,sprumont2019relative,sprumont2025two} and ours. 
\citet{sprumont2018belief} introduced the framework that we adopt and characterized belief-weighted Nashian aggregation rules. These rules evaluate each act by the weighted product of the normalized utility levels, where the weight vector is determined with reference to the profile of beliefs.  
These evaluation rules exhibit inequality aversion with respect to 0--1 normalized utility levels. 
However, as in \citet{nash1950bargaining}, this inequality-averse attitude is merely the consequence of a stronger independence axiom with respect to changes in problems, which requires invariance even when the problem expansion is not redundant in the sense that it alters some individual's best outcome.
That is, the inequality aversion inherent in these Nashian aggregation rules is obtained as a byproduct of axioms unrelated to fairness. 
On the other hand, our theorem derives the relative fair aggregation rules by imposing a property related to ex-ante fairness: \textit{weak preference for mixing}.

\citet{sprumont2019relative}  characterized (belief-weighted) relative utilitarianism using Savage's P2, a stronger independence axiom across problems (cf.\ Footnote~\ref{fn:IIE}), and their basic axioms.
In contrast, our characterization of the relative fair aggregation rules (Theorem~\ref{thm_relativefair}) shows that replacing P2 with our two moderate axioms yields a class of aggregation rules that can address the criticisms of utilitarianism (note that we employ different independence and continuity axioms). 
Moreover, by modifying one of our new axioms, we have provided another characterization of the relative utilitarian aggregation rules studied in \citet{sprumont2019relative}.

Note that \citet{sprumont2018belief,sprumont2019relative} mainly characterized aggregation rules where the weights over the individuals can vary depending on the belief profile. 
In contrast, the weight set of a relative fair aggregation rule is independent of the belief profile. 
This difference follows from \textit{belief irrelevance}. 
In our main theorem, this axiom plays an important role in making \textit{restricted certainty independence} effective in all belief profiles, as discussed after Lemma~\ref{lem:4addCI}.

Although relative utilitarianism has been examined in the context of preference aggregation (e.g., \citet{dhillon1999relative}), rules incorporating normalization and inequality aversion have, to the best of our knowledge, been studied only by \citet{sprumont2013relative}. 
Sprumont characterized the relative leximin aggregation rule in risky situations. 
In an earlier version of this paper, we also provided an axiomatic foundation for the relative leximin aggregation rule (\citet{kurata2026collective}).

Finally, we mention \citet{sprumont2025two}, in which the aggregation of time preferences was studied. 
Sprumont considered a fixed set of feasible outcomes and introduced a new invariance axiom with respect to order-preserving functions applied to outcomes and utility functions. 
Together with standard axioms, such as the  Pareto principle and time consistency, the invariance axiom characterizes the 0--1 normalization.
Since our setup fundamentally differs from that of \citet{sprumont2025two}, our result cannot be directly applied to Sprumont's framework. 
Providing axiomatic foundations for the counterparts of relative fair aggregation rules in a dynamic setup would be a promising direction for future research. 

\subsection{Rules with 0--1 normalization} \label{sec6:normalization}

Rules with the 0--1 normalization have been widely studied. 
\citet{karni1998impartiality}, \citet{dhillon1999relative}, and \citet{segal2000let} provided axiomatic foundations for relative utilitarianism in the context of preference aggregation.
For relatively recent works, see also \cite{borgers2017revealed}, \citet{marchant2018wp}, \citet{sprumont2019relative,sprumont2025two}, \citet{brandl2021belief}, and \citet{karni2024impartiality}. 

It should be mentioned that many papers on axiomatic bargaining theory have studied rules involving the 0--1 normalization. 
For instance, \citet{pivato2009twofold} and \citet{baris2018timing} characterized the relative utilitarian solution. 
While rules with egalitarian concerns and the 0--1 normalization have rarely been examined in the context of preference aggregation, they have played a central role in axiomatic bargaining theory. 
A solution concept that incorporates the 0--1 normalization and an egalitarian attitude was proposed by \citet{kalai1975other}. Their solution chooses the weakly Pareto optimal outcome that is proportional to the maximum utility levels that individuals can achieve. 
This solution can be interpreted as a choice rule that first normalizes individual utilities into the 0--1 interval and then chooses an outcome in an egalitarian way. 
Under this interpretation, the relative fair aggregation rules are similar to bargaining solutions, such as the Kalai--Smorodinsky solution. 
Indeed, \citet{nakamura2026social} has considered bargaining solutions that correspond to the relative fair aggregation rules. 
By contrast, we have derived these rules using a mixture operation that can be naturally formalized owing to our framework with uncertainty. 

\begin{appendix}
\section{Proofs}\label{app}

In each characterization result, we prove only the sufficiency of the axioms because their necessity is straightforward to show. 

\subsection{Proof of Lemma~\ref{lem:basic}}
Let $\mathbf{R}$ be an aggregation rule that satisfies \textit{weak Pareto principle}, \textit{event continuity}, \textit{belief irrelevance}, and \textit{independence of redundant outcomes}. 

Let $\tilde{X} = \{\tilde{x}_M\}_{M\in 2^N} \subset \mathbb{X}$ be such that $|\tilde{X}| = 2^{|N|}$. 
Take a preference profile $\tilde{R}_N$ on $\tilde{X}$ such that for each $M\in 2^N$ and each $i\in N$, $U^*_i (\tilde{x}_M; \tilde{X}, \tilde{R}_N) = 1$ if $i\in M$, and $U^*_i (\tilde{x}_M; \tilde{X}, \tilde{R}_N) = 0$ otherwise.
By Lyapunov's convexity theorem, for each $\mathbf{u} \in [0,1]^N$, there exists $f\in F_{\tilde{X}}$ such that  $U^* (f; \tilde{X}, \tilde{R}_N) = \mathbf{u}$.%
\footnote{
To see this, note that there exists a vector $( \alpha_M)_{M\in 2^N}$ of nonnegative real numbers such that $\sum_{M\in 2^N} \alpha_M = 1$ and $\mathbf{u} = \sum_{M\in 2^N} \alpha_M U^* (\tilde{x}_M; \tilde{X}, \tilde{R}_N)$ because $\{ U^* (\tilde{x}_M; \tilde{X}, \tilde{R}_N) \}_{M\in 2^N}$ is the set of extreme points of $[0,1]^N$.
By Lyapunov's convexity theorem, there exists a partition $\{ E_M \}_{M\in 2^N}$ of $\Omega$ such that $p_i (E_M) = \alpha_M$ for all $M\in 2^N$ and all $i\in N$. 
Let $f\in F_{\tilde{X}}$ be such that $f(E_M) = \{\tilde{x}_M\}$ for all $M\in 2^N$. 
By construction, $U^* (f; \tilde{X}, \tilde{R}_N) = \sum_{M\in 2^N} \alpha_M U^* (\tilde{x}_M; \tilde{X}, \tilde{R}_N) = \mathbf{u}$ holds.
}
Define the binary relation $\succsim$ over $[0,1]^N$ by the condition that for all $\mathbf{u}, \mathbf{v} \in [0,1]^N$, $\mathbf{u} \succsim \mathbf{v} $ if and only if there exist $f,g \in F_{\tilde{X}}$ such that $U^* (f; \tilde{X}, \tilde{R}_N) = \mathbf{u}$, $U^* (g; \tilde{X}, \tilde{R}_N) = \mathbf{v}$, and $f\, \mathbf{R} (\tilde{X}, \tilde{R}_N) \, g$.
By \textit{weak Pareto principle}, this relation is well-defined. 
Since $\mathbf{R} (\tilde{X}, \tilde{R}_N)$ is complete and transitive, so is $\succsim$. 
The symmetric and asymmetric parts are denoted by $\sim$ and $\succ$, respectively. 
The following claim shows that the labeling of outcomes in $\tilde{X}$ does not matter.

\begin{claim}
\label{claim:retake}
    Let $X' = \{x'_M\}_{M\in 2^N} \subset \mathbb{X} \backslash \tilde{X}$, 
    and define $R'_N$ on $X'$ by the condition that $p^* (R'_N) = p^* (\tilde{R}_N)$ and $U^* (x'_M; X', R'_N) = U^* (\tilde{x}_M; \tilde{X}, \tilde{R}_N)$ for all $M\in 2^N$. 
    Then, for all $f', g'\in F_{X'}$, 
    \begin{equation}
        f' \, \mathbf{R} (X', R'_N) \, g' \iff U^* (f'; X', R'_N) \succsim U^* (g'; X', R'_N). 
    \end{equation} 
\end{claim}

\begin{proof}\renewcommand{\qedsymbol}{$||$}
Assume that for some $f', g'\in F_{X'}$,  $f' \, \mathbf{R} (X', R'_N) \, g'$ and $U^* (g' ; X', R'_N) \succ U^* (f'; X', R'_N)$. 
By the definition of $\succsim$, there exist $\tilde{f},\tilde{g} \in F_{\tilde{X}}$ such that $U^* (\tilde{f}; \tilde{X}, \tilde{R}_N) = U^* (f' ; X', R'_N)$, $U^* (\tilde{g}; \tilde{X}, \tilde{R}_N) = U^* (g' ; X', R'_N)$, and $\tilde{g} \, \mathbf{P} (\tilde{X}, \tilde{R}_N) \, \tilde{f}$. 

Let $(X^\cup, R^\cup_N) \in \mathcal{D}$ be such that (i) $X^\cup = \tilde{X} \cup X'$, (ii) for all $x\in X^\cup$, $U^\ast (x; X^\cup, R^\cup_N) = U^\ast (x; \tilde{X}, \tilde{R}_N)$ if $x\in \tilde{X}$, and $U^\ast (x; X^\cup, R^\cup_N) = U^\ast (x; X', R'_N)$ otherwise, and 
(iii) $p^* (R^\cup_N) = p^* (\tilde{R}_N)$.
Then, since $(X^\cup, R^\cup_N)$ is a redundant-outcome expansion of $(\tilde{X}, \tilde{R}_N)$, 
\textit{independence of redundant outcomes} implies $\tilde{g} \, \mathbf{P} (X^\cup, R^\cup_N) \, \tilde{f}$. 
Since  $(X^\cup, R^\cup_N)$ is also a redundant-outcome expansion of $(X', R'_N)$, 
\textit{independence of redundant outcomes} implies $f' \, \mathbf{R} (X^\cup, R^\cup_N) \, g'$. 
By construction, $U^* (\tilde{f} ; X^\cup, R^\cup_N) = U^* ( f' ; X^\cup, R^\cup_N)$ and  $U^* (\tilde{g} ; X^\cup, R^\cup_N) = U^* ( g' ; X^\cup, R^\cup_N)$. 
Then, \textit{weak Pareto principle} implies $\tilde{f} \, \mathbf{I} (X^\cup, R^\cup_N) \, f'$ and $\tilde{g} \, \mathbf{I} (X^\cup, R^\cup_N) \, g'$, which is a contradiction to the transitivity of $\succsim$.
To prove the converse, assume that for some $f'', g''\in F_{X'}$,  $g'' \, \mathbf{P} (X', R'_N) \, f''$ and $U^* (f'' ; X', R'_N) \succsim U^* (g''; X', R'_N)$. 
Then, by a similar argument, we can see that \textit{weak Pareto principle} and \textit{independence of redundant outcomes} yield a contradiction to the transitivity of $\succsim$.  
\end{proof}

\begin{claim}
    For all $\mathbf{u}, \mathbf{v} \in [0,1]^N$ with $\mathbf{u} \succsim \mathbf{v} $ and all $(X, R_N) \in \mathcal{D}$, if there exist $f, g \in F_X$ such that $U^* (f; X, R_N) = \mathbf{u}$, $U^* (g ; X, R_N) = \mathbf{v}$, then $f\, \mathbf{R} (X, R_N) \, g$. 
\end{claim}

\begin{proof}\renewcommand{\qedsymbol}{$||$}
Suppose to the contrary that there exist $\mathbf{u}, \mathbf{v} \in [0,1]^N$, $(X, R_N) \in \mathcal{D}$, and $f_\mathbf{u} ,f_\mathbf{v}\in F_X$ 
such that $\mathbf{u} \succsim \mathbf{v} $,  $U^* (f_\mathbf{u}; X, R_N) = \mathbf{u}$, $U^* (f_\mathbf{v} ; X, R_N) = \mathbf{v}$, and $f_\mathbf{v} \, \mathbf{P} (X, R_N) \, f_\mathbf{u}$. 
By Claim~\ref{claim:retake}, we can assume without loss of generality that $\tilde{X} \cap X = \emptyset$.

Take $x_\mathbf{u}, x_\mathbf{v} \in \mathbb{X} \backslash (\tilde{X} \cup X)$, and let $X^1 = X\cup \{x_\mathbf{u}, x_\mathbf{v} \}$. 
Let $R^1_N$ be a preference profile on $X^1 $ such that 
(i) for all $x\in X$, $U^*(x; X^1, R^1_N) =U^*(x; X, R_N)$,
(ii) $ U^*(x_\mathbf{u}; X^1, R^1_N) = \mathbf{u}$ and  $ U^*(x_\mathbf{v}; X^1, R^1_N) = \mathbf{v}$, 
and
(iii) $p^* (R_N^1) = p^* (R_N)$. 
Then, since $(X^1, R^1_N)$ is a redundant-outcome expansion of $(X, R_N)$, \textit{independence of redundant outcomes} implies $f_\mathbf{v} \, \mathbf{P} (X^1, R^1_N) \, f_\mathbf{u}$.
By \textit{weak Pareto principle} and the transitivity of $\mathbf{R} (X^1, R^1_N)$, we have $x_\mathbf{v} \, \mathbf{P} (X^1, R^1_N) \, x_\mathbf{u}$.

Take a belief profile $p_N$ such that there exists a partition $\{ E_k\}_{k\in N}$ of $\Omega$ with $p_i (E_i) =1$ for each $i\in N$. 
Let $R^{1+}_N$ be the preference profile on $X^1 $ such that 
(i) for all $x\in X^1$, $U^*(x; X^1, R^{1+}_N) = U^*(x; X^1, R^1_N)$ 
and 
(ii) $p^* (R^{1+}_N) = p_N $. 
By \textit{belief irrelevance}, $x_\mathbf{v} \, \mathbf{P} (X^1, R^{1+}_N) \, x_\mathbf{u}$ holds.

Next, take a set $\{x_M\}_{M\in 2^N} \subset \mathbb{X} \backslash (\tilde{X} \cup X^1)$ with $2^{|N|}$ distinct outcomes, and let $X^2 = X^1\cup \{x_M\}_{M\in 2^N}$. 
Let $R_N^{2-}$ be a preference profile over $X^2$ such that (i) $ U^* (f; X^2, R_N^{2-}) = U^* (f; X^1,R_N^{1+})$ for all $f\in F_{X^1}$, (ii) for each $M\in 2^N$ and each $i\in N$, $U^*_i (x_M; X^2, R_N^{2-}) = 1$ if $i\in M$, and $U^*_i (x_M; X^2, R_N^{2-}) = 0$ otherwise, and (iii) $p^* (R^{2-}_N) = p_N $. 
Note that $(X^2, R_N^{2-})$ is a redundant-outcome expansion of $(X^1, R_N^{1+})$. 
To see this, for each $i\in N$, write $x_i^+ \in X^1$ (resp.\ $x_i^- \in X^1$) for an outcome such that $U_i^*(x_i^+; X^1, R^{1+}_N) = 1$ (resp.\ $U_i^*(x_i^-; X^1, R^{1+}_N) = 0$).
For each $M\in 2^N$, let $f_M \in F_{X^1}$ be such that for each $i\in N$, $f_M (E_i) = \{ x_i^+ \}$ if $i\in M$, and $f_M (E_i) = \{ x_i^- \}$ otherwise. 
Then, by the definition of $\{ E_k\}_{k\in N}$, we have $U^* (x_M ; X^2, R_N^{2-}) = U^* (f_M ; X^1, R_N^{1+}) $, as required. 
By \textit{independence of redundant outcomes}, we have $x_\mathbf{v} \, \mathbf{P} (X^2, R^{2-}_N) \, x_\mathbf{u}$.
Define the preference profile $R_N^2$ over $X^2$ by the condition that 
(i) for all $x\in X^2$, $U^*(x; X^2, R^2_N) = U^*(x; X^2, R^{2-}_N)$ 
and 
(ii) $p^* (R^2_N) = p^* (\tilde{R}_N)$. 
By \textit{belief irrelevance}, $x_\mathbf{v} \, \mathbf{P} (X^2, R^2_N) \, x_\mathbf{u}$ holds.

Let $\tilde{f}_\mathbf{u}, \tilde{f}_\mathbf{v} \in F_{\tilde{X}}$ be such that $U^*(\tilde{f}_\mathbf{u}; \tilde{X}, \tilde{R}_N) = \mathbf{u}$ and $U^*(\tilde{f}_\mathbf{v}; \tilde{X}, \tilde{R}_N) = \mathbf{v}$.
By $\mathbf{u} \succsim \mathbf{v} $, we have $\tilde{f}_\mathbf{u}\, \mathbf{R} (\tilde{X}, \tilde{R}_N) \,  \tilde{f}_\mathbf{v}$. 
Take $\tilde{x}_\mathbf{u}, \tilde{x}_\mathbf{v} \in \mathbb{X} \backslash (\tilde{X} \cup X^2)$, and let $\tilde{X}^1 = \tilde{X} \cup \{ \tilde{x}_\mathbf{u}, \tilde{x}_\mathbf{v} \}$. 
Let $\tilde{R}^1_N$ be the preference profile on $\tilde{X}^1 $ such that 
(i) for all $x\in \tilde{X}$, $ U^*(x; \tilde{X}^1, \tilde{R}^1_N) = U^*(x; \tilde{X}, \tilde{R}_N)$,
(ii) $ U^*(\tilde{x}_\mathbf{u}; \tilde{X}^1, \tilde{R}^1_N) = \mathbf{u}$ and  $ U^*(\tilde{x}_\mathbf{v}; \tilde{X}^1, \tilde{R}^1_N) = \mathbf{v}$, 
and
(iii) $p^* (\tilde{R}_N^1) = p^* (\tilde{R}_N)$. 
Since $(\tilde{X}^1, \tilde{R}^1_N)$ is a redundant-outcome expansion of $(\tilde{X}, \tilde{R}_N)$, \textit{independence of redundant outcomes} implies $\tilde{f}_\mathbf{u}\, \mathbf{R} (\tilde{X}^1, \tilde{R}^1_N) \,  \tilde{f}_\mathbf{v}$. 
By \textit{weak Pareto principle} and the transitivity of $\mathbf{R} (\tilde{X}^1, \tilde{R}^1_N)$, we have $\tilde{x}_\mathbf{u}\, \mathbf{R} (\tilde{X}^1, \tilde{R}^1_N) \,  \tilde{x}_\mathbf{v}$. 

Finally, we construct the problem $(X^\cup, R^\cup_N)$ such that (i) $X^\cup = X^2 \cup \tilde{X}^1$, (ii) for all $x\in X^\cup$, $U^\ast (x; X^\cup, R^\cup_N) = U^\ast (x; X^2, R^2_N)$ if $x\in X^2$, and $U^\ast (x; X^\cup, R^\cup_N) = U^\ast (x; \tilde{X}^1, \tilde{R}^1_N)$ otherwise, and (iii) $p^* (R_N^\cup) = p^* (\tilde{R}_N)$. 
Since $(X^\cup, R^\cup_N)$ is a redundant-outcome expansion of $(\tilde{X}^1, \tilde{R}^1_N)$, \textit{independence of redundant outcomes} implies $\tilde{x}_\mathbf{u}\, \mathbf{R} (X^\cup, R^\cup_N) \,  \tilde{x}_\mathbf{v}$. 
Also, since $(X^\cup, R^\cup_N)$ is a redundant-outcome expansion of $(X^2, R^2_N)$, \textit{independence of redundant outcomes} implies 
$x_\mathbf{v} \, \mathbf{P} (X^\cup, R^\cup_N) \, x_\mathbf{u}$. 
By construction, $U^*(x_\mathbf{u}; X^\cup, R^\cup_N) = U^*(\tilde{x}_\mathbf{u}; X^\cup, R^\cup_N) $ and  $U^*(x_\mathbf{v}; X^\cup, R^\cup_N) = U^*(\tilde{x}_\mathbf{v}; X^\cup, R^\cup_N) $.
\textit{Weak Pareto principle} implies that $x_\mathbf{u} \, \mathbf{I} (X^\cup, R^\cup_N) \, \tilde{x}_\mathbf{u}$ and $x_\mathbf{v} \, \mathbf{I} (X^\cup, R^\cup_N) \, \tilde{x}_\mathbf{v}$, 
which is a contradiction to the transitivity of $\succsim$. 
\end{proof}

Similarly, we can prove the following claim.

\begin{claim}
    For all $\mathbf{u}, \mathbf{v} \in [0,1]^N$ with $\mathbf{u} \succ \mathbf{v} $ and all $(X, R_N) \in \mathcal{D}$, if there exist $f, g \in F_{X}$ such that $U^* (f; X, R_N) = \mathbf{u}$, $U^* (g; X, R_N) = \mathbf{v}$, then $f\, \mathbf{P} (X, R_N) \, g$.
\end{claim}

Therefore, for all $(X, R_N)\in \mathcal{D}$ and all $f,g\in F_X$, $f\, \mathbf{R} (X, R_N) \, g$ if and only if $U^*(f; X, R_N) \succsim U^*(g; X, R_N)$. 
By \textit{weak Pareto principle}, $\succsim$ is monotonic, that is, for all $\mathbf{u}, \mathbf{v} \in [0,1]^N$,  $\mathbf{u} \geq \mathbf{v}$ implies $\mathbf{u} \succsim \mathbf{v}$, and $\mathbf{u} \gg \mathbf{v}$ implies $\mathbf{u} \succ \mathbf{v}$.

\begin{claim}
    For all $\mathbf{v} \in [0,1]^N$,  the sets $\{ \mathbf{u} \in [0,1]^N \mid \mathbf{u} \succ \mathbf{v}\}$  and $\{ \mathbf{u} \in [0,1]^N \mid  \mathbf{v} \succ \mathbf{u} \}$ are open. 
\end{claim}

\begin{proof}\renewcommand{\qedsymbol}{$||$}
Let $\mathbf{u}, \mathbf{v} \in [0,1]^N$ be such that  $\mathbf{u} \succ\mathbf{v}$. 
Note that $\mathbf{u} \neq \mathbf{0}$. 
Let $(X, R_N) \in\mathcal{D}$ be such that (i) there exist $x_\mathbf{u}, x_\mathbf{v}, x_\mathbf{0} \in X$ with $U^* (x_\mathbf{u}; X, R_N) = \mathbf{u}$, $U^* (x_\mathbf{v}; X, R_N) = \mathbf{v}$, and $U^* (x_\mathbf{0}; X, R_N) = \mathbf{0}$, and (ii) $p^* (R_i) =p^* (R_j)$ for all $i,j\in N$. 
By $\mathbf{u} \succ\mathbf{v}$, we have $x_\mathbf{u} \, \mathbf{P}(X, R_N ) \,x_\mathbf{v} $. 
By \textit{event continuity}, there exists a partition $\{ E_k \}_{k=1}^m$ such that for all $k \in \{1,2,\ldots, m\}$, $x_\mathbf{0} E_k x_\mathbf{u} \, \mathbf{P}(X, R_N ) \,x_\mathbf{v} $. 
By the construction of $p^* (R_N) $, we can take $E \in \{ E_k \}_{k=1}^m$ such that $p_i (E) > 0$ for all $i\in N$. 
Let $\mathbf{w} \in [0,1]^N$ be such that $\mathbf{w} = U^* (x_\mathbf{0} E x_\mathbf{u}; X, R_N) = p^* (R_i) (E)  \mathbf{0} + (1-p^* (R_i)(E)) \mathbf{u}$. 
For each $i\in N$, $\mathbf{w}_i < \mathbf{u}_i$ if $\mathbf{u}_i > 0$, and  $\mathbf{w}_i = \mathbf{u}_i =0$ otherwise. 
By $x_\mathbf{0} E x_\mathbf{u} \, \mathbf{P}(X, R_N ) \,x_\mathbf{v} $, we have $\mathbf{w} \succ \mathbf{v}$. 

Let $A \subset \{ \mathbf{u}' \in \mathbb{R}^N \mid \mathbf{u}' \geq    \mathbf{w} \}$ be a nonempty set such that $A$ is open relative to $[0,1]^N$ and includes $\mathbf{u}$. 
Note that by $\mathbf{w} \succ \mathbf{v}$ and the monotonicity of $\succsim$, $\mathbf{u}' \succ \mathbf{v}$ for all $\mathbf{u}' \in A$. 
Therefore,  $\{ \mathbf{u}' \in [0,1]^N \mid \mathbf{u}' \succ \mathbf{v}\}$ is open. 
By a similar argument, we can show that  $\{ \mathbf{u}' \in [0,1]^N \mid  \mathbf{v} \succ  \mathbf{u}'\}$ is also open. 
\end{proof}

Therefore, there exists a monotonic continuous function $\psi: [0,1]^N \to \mathbb{R}$ such that for all $\mathbf{u}, \mathbf{v}\in[0,1]^N$, $\mathbf{u} \succsim \mathbf{v}$ if and only if $\psi( \mathbf{u}) \geq \psi(\mathbf{v})$. 
\qed

\subsection{Proof of Lemma~\ref{lem:quasiconcav}}
Let $\mathbf{R}$ be an aggregation rule that satisfies \textit{weak Pareto principle}, \textit{event continuity}, \textit{belief irrelevance}, \textit{independence of redundant outcomes}, and \textit{weak preference for mixing}. 
Define the binary relation $\succsim$ over $[0,1]^N$ by the condition that, for all $\mathbf{u}, \mathbf{v}\in [0, 1]^N$, $\mathbf{u} \succsim \mathbf{v}$ if there exist $(X,R_N) \in \mathcal{D}$ and $f, g \in F_X$ such that  $\mathbf{u} = U^\ast(f;X, R_N)$, $\mathbf{v} = U^\ast(g;X, R_N)$, and  $f \, \mathbf{R} (X, R_N) \, g$. 
By Lemma~\ref{lem:basic}, this binary relation is well-defined. The symmetric and asymmetric parts are denoted by $\sim$ and $\succ$, respectively.

First, we show that for all $\mathbf{u}, \mathbf{v} \in [0, 1]^N$, $\mathbf{u} \sim \mathbf{v}$ implies $\frac{1}{2}(\mathbf{u} + \mathbf{v}) \succsim \mathbf{u}$. 
Take $\mathbf{u}, \mathbf{v} \in [0, 1]^N$ such that $\mathbf{u} \sim \mathbf{v}$. 
By the definition of $\succsim$ and Lemma~\ref{lem:basic}, there exist $(X, R_N) \in \mathcal{D}$ and constant acts $x, y \in X$ such that $U^*(x; X, R_N) = \mathbf{u}$, $U^*(y; X, R_N) = \mathbf{v}$, and $x \,\mathbf{I}(X, R_N)\, y$. 
By Lyapunov's convexity theorem, there exists $E \subset \Omega$ such that $p_i(E) = \frac{1}{2}$ for all $i \in N$. 
It follows from \textit{weak preference for mixing} that $x E y \,\mathbf{R}(X, R_N)\, x$.
Since $R_i$ is an SEU preference of individual $i \in N$, we have $U^*(xEy; X, R_N) = \frac{1}{2}(\mathbf{u} + \mathbf{v})$.  
Therefore,  $\frac{1}{2}(\mathbf{u} + \mathbf{v}) \succsim \mathbf{u} $ holds.

Next, we show that for all $\mathbf{u} \in [0, 1]^N$, the set $ \{\mathbf{v} \in [0, 1]^N \mid \mathbf{v} \succsim \mathbf{u}\}$ is convex.
Suppose to the contrary that there exists $\mathbf{u} \in [0, 1]^N$ such that the set $ \{\mathbf{v} \in [0, 1]^N \mid \mathbf{v} \succsim \mathbf{u}\}$ is not convex. 
Then, there exist $\mathbf{v}^1, \mathbf{v}^2 \in [0,1]^N$ such that $\mathbf{v}^1 \sim \mathbf{v}^2 \succsim \mathbf{u} \succ \frac{1}{2}\mathbf{v}^1 + \frac{1}{2}\mathbf{v}^2$.%
\footnote{
Note that by the argument in the proof of Lemma~\ref{lem:basic}, the sets $ \{\mathbf{v} \in [0, 1]^N \mid \mathbf{v} \succsim \mathbf{u}\}$ and $ \{\mathbf{v} \in [0, 1]^N \mid \mathbf{u} \succsim \mathbf{v}\}$ are closed. Hence, we can take such vectors without loss of generality.
}
This contradicts the result in the previous paragraph. \qed

\subsection{Proof of Proposition~\ref{prop:pmact}}

Let $(X, R_N) \in \mathcal{D}$, $f, g \in F_X$, and $\alpha\in (0,1)$. 
Take any $y \in f(\Omega)$ and any $y' \in g(\Omega)$.
For notational simplicity, in this proof, let $p_N = (p_1, p_2, \ldots, p_n)$ denote $p^\ast (R_N)$.
By Lyapunov's convexity theorem, the set 
\begin{equation*}
    \mathbb{P}_{(y, y')} = \Big\{ \big( p_1(E), p_2(E), \ldots, p_n (E) \big)  ~ \Big| ~E \subset f^{-1} (y) \cap g^{-1} (y')  \Big\}
\end{equation*}
is a convex subset of $[0,1]^N$. 
Note that since $\emptyset$ and $f^{-1}(y) \cap  g^{-1} (y')$ are subsets of $f^{-1}(y) \cap  g^{-1} (y')$, two elements $\mathbf{0}$ and $ \Big(p_1 \big( f^{-1} (y) \cap g^{-1} (y') \big), \ldots, p_n  \big( f^{-1} (y) \cap g^{-1} (y') \big)\Big) $ are in $\mathbb{P}_{(y, y')}$. By the convexity of $\mathbb{P}_{(y, y')}$, we have 
\begin{equation*}
    \alpha \left(p_1 \big( f^{-1} (y) \cap g^{-1} (y') \big) , \ldots, p_n  \big( f^{-1} (y) \cap g^{-1} (y') \big) \right) + (1- \alpha) \mathbf{0}  \in \mathbb{P}_{(y, y')}.
\end{equation*}
Therefore, for each  $(y, y') \in f(\Omega) \times g(\Omega)$, there exists an event $E_{(y, y')} \subset  f^{-1} (y) \cap g^{-1} (y')$ such that $p_i  \big(E_{(y, y')} \big) = \alpha p_i \big(  f^{-1} (y) \cap g^{-1} (y') \big)$ for all $i \in N$. 

Let $E^\ast = \bigcup_{(y,y')\in f(\Omega) \times g(\Omega)} E_{(y, y')} $.\footnote{Note that $E^\ast$ depends on $f$, $g$, and $\alpha$.} 
Since $E_{(y, y')}$ and $E_{(z, z')}$ are disjoint for all distinct $(y, y'),(z, z') \in f(\Omega) \times g(\Omega)$,
we have $p_i (E^\ast ) = \sum_{(y,y')\in f(\Omega) \times g(\Omega)} \alpha  p_i \big(f^{-1} (y) \cap g^{-1} (y')\big)= \alpha$ for all $i\in N$. 
Define $f_{\alpha}g \in F_X$ by, for all $\omega\in \Omega$,
$(f_{\alpha} g) (\omega) = f(\omega)$ if $\omega\in E^\ast$, and $(f_{\alpha} g) (\omega) = g(\omega)$ otherwise.
By construction, for all $i\in N$ and all $x\in X$, 
\begin{equation*}
    p_i \big((f_{\alpha} g)^{-1} (x)\big) = \alpha p_i \big(f^{-1} (x)\big) + (1-\alpha) p_i \big(g^{-1} (x)\big).  \tag*{\qed}
\end{equation*}

\subsection{Proof of Lemma~\ref{lem:4addCI}}

Let $\mathbf{R}$ be an aggregation rule that satisfies \textit{weak Pareto principle}, \textit{event continuity}, \textit{belief irrelevance}, \textit{independence of redundant outcomes}, and
\textit{restricted certainty independence}. 
Let $\psi :[0,1]^N \rightarrow \mathbb{R}$ be the monotonic continuous function in Lemma~\ref{lem:basic}. 
Without loss of generality, we assume that for all $c \in [0,1]$, $\psi (c\mathbf{1}) = c$. 
Define the binary relation $\succsim$ over $[0,1]^N$ as in the proof of Lemma~\ref{lem:quasiconcav}. 

Let $c\in [0,1]$ and $\mathbf{u}, \mathbf{v}\in [0, 1]^N$. 
First, we prove that for all $\alpha \in (0,1)$, $\mathbf{u}\succsim \mathbf{v}$ if and only if $\alpha\mathbf{u} + (1- \alpha) c\mathbf{1} \succsim \alpha \mathbf{v} + (1- \alpha) c\mathbf{1}$. 
Suppose $\mathbf{u}\succsim \mathbf{v}$. 
Take $( X, R_N ) \in \mathcal{D}$ and $f, g \in F_X$ such that (i)  $u_i^\ast (\cdot;  X, R_N) = u_j^\ast (\cdot;  X, R_N)$ for each $i,j\in N$, (ii) for some $x \in X$, $c \mathbf{1} = u^\ast(x;   X, R_N)$, (iii) there exist  $x^\ast,  x_\ast \in X$ with $x^\ast R_i y R_i x_\ast$ for all $i\in N$ and all $y\in X$, and (iv)
\begin{align*}
    p^\ast(R_N) (f^{-1}(x^\ast)) &= \mathbf{u}, ~~~~p^\ast(R_N) (f^{-1}(x_\ast)) =\mathbf{1} - \mathbf{u}, \\
    p^\ast(R_N) (g^{-1}(x^\ast)) &= \mathbf{v}, ~~~~p^\ast(R_N) (g^{-1}(x_\ast)) =\mathbf{1} -  \mathbf{v}.
\end{align*}
Since  $R_i$ is an SEU preference for each $i\in N$, (iii) and (iv) imply that  $U^\ast (f;  X, R_N) =  \mathbf{u}$ and $U^\ast (g;  X, R_N) =  \mathbf{v}$. 
By $\mathbf{u}\succsim \mathbf{v}$ and the construction of $\succsim$, $f \, \mathbf{R}(X, R_N) \, g$. 
By \textit{restricted certainty independence}, for any  $\alpha \in (0,1)$, $f \, \mathbf{R}(X, R_N) \, g $ if and only if $f_{\alpha} x \, \mathbf{R}(X, R_N ) \, g_{\alpha} x$. 
Since  $R_i$ is an SEU preference for each $i\in N$, 
$U^\ast (f_{\alpha} x;  X, R_N) = \alpha \mathbf{u} + (1- \alpha) c\mathbf{1}$ and  $U^\ast (g_{\alpha} x; X, R_N) = \alpha \mathbf{v} + (1- \alpha) c\mathbf{1}$. 
Therefore, for all $\alpha \in (0,1)$, 
\begin{equation}
\label{eq_constrainedCI}
    \mathbf{u}\succsim \mathbf{v} \iff \alpha \mathbf{u} + (1- \alpha) c\mathbf{1} \succsim \alpha \mathbf{v} + (1- \alpha) c\mathbf{1}.
\end{equation}

We then prove that for all $\mathbf{u},\mathbf{v}\in [0,1]^N$ and all $\alpha \in \mathbb{R}_{++}$ such that $\alpha \mathbf{u} , \alpha \mathbf{v}  \in [0,1]^N$, it holds that
$\mathbf{u}\succsim \mathbf{v}$ if and only if $\alpha \mathbf{u}  \succsim \alpha \mathbf{v}$. If $\alpha \in (0,1)$, then we can prove the above by setting $c = 0$ in \eqref{eq_constrainedCI}. 
If $\alpha > 1$, i.e., $0 < {1 \over \alpha} < 1$, then by applying \eqref{eq_constrainedCI} with $c=0$, we have
\begin{equation*}
    \mathbf{u} \succsim \mathbf{v} \iff 
    {1 \over \alpha} \alpha \mathbf{u} \succsim {1 \over \alpha} \alpha \mathbf{v}
    \iff 
    \alpha \mathbf{u} \succsim 
    \alpha \mathbf{v}. 
\end{equation*}

Therefore, for all $\mathbf{u}\in [0,1]^N$ and  all $\alpha \in \mathbb{R}_{++}$ such that $\alpha \mathbf{u} \in [0,1]^N$, since $\mathbf{u}\sim \psi (\mathbf{u})\mathbf{1} $ is equivalent to $\alpha \mathbf{u}   \sim \alpha \psi (\mathbf{u})\mathbf{1}$, we have $\psi (\alpha \mathbf{u} ) = \psi (\alpha \psi (\mathbf{u})\mathbf{1}) = \alpha \psi (\mathbf{u})$, where the second equality follows from the assumption that $\psi (c\mathbf{1}) = c$ for all $c \in [0,1]$. That is, the function $\psi$ is homogeneous.

Let  $\mathbf{u},\mathbf{v}\in [0,1]^N$ and $c\in \mathbb{R}$ such that $ \mathbf{u} +  c \mathbf{1} ,  \mathbf{v} +  c \mathbf{1} \in [0,1]^N $. Then, $-1 \le c \le 1$ holds. If $c \in [0,1]$, then by applying \eqref{eq_constrainedCI} twice,
\begin{align*}
    \mathbf{u}\succsim \mathbf{v}
    &\iff 
    {1 \over 2} \mathbf{u} + {1\over 2} ( c + 0) \mathbf{1} \succsim {1 \over 2} \mathbf{v} + {1\over 2} ( c + 0) \mathbf{1} \\
    &\iff {1 \over 2} ( \mathbf{u}  + c \mathbf{1})+ {1 \over 2}  \mathbf{0} \succsim {1 \over 2}   (\mathbf{v}+ c  \mathbf{1} ) + {1 \over 2}   \mathbf{0} \\
    &\iff  \mathbf{u}  + c \mathbf{1} \succsim  \mathbf{v}  + c \mathbf{1}. 
\end{align*}
On the other hand, if  $c \in [-1,0]$, then by applying \eqref{eq_constrainedCI} twice,
\begin{align*}
    \mathbf{u}\succsim \mathbf{v}
    &\iff 
    {1 \over 2} \mathbf{u} \succsim {1 \over 2} \mathbf{v} \\
    &\iff {1 \over 2} ( \mathbf{u}  + c \mathbf{1})+ {-c \over 2}  \mathbf{1} \succsim {1 \over 2}   (\mathbf{v}+ c  \mathbf{1} ) + {-c \over 2}  \mathbf{1} \\
    &\iff  \mathbf{u}  + c \mathbf{1} \succsim  \mathbf{v}  + c \mathbf{1}. 
\end{align*}
Thus, for all $\mathbf{u}\in [0,1]^N$ and all $c \in \mathbb{R}$ such that $\mathbf{u} +  c \mathbf{1} \in [0,1]^N$, we have that $\mathbf{u} \sim \psi (\mathbf{u}) \mathbf{1} $ is equivalent to $ \mathbf{u} +  c \mathbf{1} \sim \psi (\mathbf{u}) \mathbf{1} +  c \mathbf{1}$. 
By the definition of $\psi$,  $\psi ( \mathbf{u} +  c \mathbf{1}) = \psi ( \psi (\mathbf{u}) \mathbf{1} +  c \mathbf{1}) =\psi (\mathbf{u})  +  c$, where the second equality follows from the assumption that $\psi (c\mathbf{1}) = c$ for all $c \in [0,1]$. That is, the function $\psi$ is translation-invariant. \qed

\subsection{Proof of Theorem~\ref{thm_relativefair}} \label{proof:thm1}

Let $\mathbf{R}$ be an aggregation rule that satisfies 
\textit{weak Pareto principle}, \textit{event continuity}, \textit{belief irrelevance}, \textit{independence of redundant outcomes}, \textit{weak preference for mixing}, and \textit{restricted certainty independence}. 
By Lemmas \ref{lem:quasiconcav} and \ref{lem:4addCI}, 
there exists a monotonic, continuous, quasiconcave, homogeneous, and translation-invariant function $\psi :[0,1]^N \rightarrow \mathbb{R}$ such that
for each $(X, R_N) \in \mathcal{D}$, $\mathbf{R} (X, R_N)$ is represented by the function $W_{(X, R_N)} :F_X \to \mathbb{R}$ defined by $W_{(X, R_N)} (f) = \psi \big(U^\ast (f;  X, R_N)\big)$ for all $f\in F_X$.

Since $\psi$ is homogeneous and translation-invariant, 
it is straightforward to prove that we can uniquely extend $\psi$ defined on $[0,1]^N$ to $\widetilde{\psi}$ defined on $\mathbb{R}^N$  such that $\widetilde{\psi}$ is a monotonic, continuous,  quasiconcave, homogeneous, and translation-invariant function. 

Let  $\text{UC}_\mathbf{0} = \big\{ \mathbf{u}\in \mathbb{R}^N ~ \big| ~ \widetilde{\psi} (\mathbf{u}) \geq \widetilde{\psi} (\mathbf{0}) \big\}$. 
By the continuity and quasiconcavity of $\widetilde{\psi}$, $\text{UC}_\mathbf{0}$ is closed and convex. 
Since the homogeneity of $\widetilde{\psi}$ implies that  $\alpha \mathbf{u}\in \text{UC}_\mathbf{0}$ for all $\mathbf{u}  \in \text{UC}_\mathbf{0}$ and all $\alpha \in \mathbb{R}_{++}$, the set $\text{UC}_\mathbf{0}$ is a nonempty closed convex cone. 

By applying the supporting hyperplane theorem to the pair consisting of  $\text{UC}_\mathbf{0}$ and $\mathbf{0}$, there exists $\mu^\ast \in \mathbb{R}^N$ such that for all $\mathbf{u}\in \text{UC}_\mathbf{0}$, 
\begin{equation}
\label{eq_support}
    \sum_{i\in N} \mu^\ast_i \mathbf{u}_i \geq 0. 
\end{equation}
Since $\text{UC}_\mathbf{0}$ includes $\mathbb{R}^N_+$ (by the monotonicity of $\widetilde{\psi}$), $\mu^\ast \in \mathbb{R}^N_+$. 
Thus, we can set $\mu^\ast\in \Delta_N$. 
Let $\mathcal{M}\subset \Delta_N$ be the set of vectors $\mu^\ast$ satisfying \eqref{eq_support} for all $\mathbf{u} \in \text{UC}_\mathbf{0}$.  

The set $\mathcal{M}$ is convex. 
To see this, let $\mu, \mu'\in \mathcal{M}$ and $\alpha\in(0,1)$. 
By the definition, for all $\mathbf{u} \in \text{UC}_\mathbf{0}$, 
$\sum_{i\in N} \mu_i  \mathbf{u}_i \geq 0$ and $\sum_{i\in N} \mu'_i  \mathbf{u}_i \geq 0$, 
which implies that for all $\mathbf{u} \in \text{UC}_\mathbf{0}$, 
$\sum_{i\in N} \left(\alpha \mu_i + (1 - \alpha) \mu'_i \right)  \mathbf{u}_i \geq 0$.

We claim that $\mathcal{M}$ is a closed set. 
Let $\{ \mu^k \}_{k \in \mathbb{N}} \subset \mathcal{M}$ be a sequence that converges to $\mu$. By the definition of $\mathcal{M}$, $\sum_{i\in N} \mu^k_i  \mathbf{u}_i \geq 0$ holds for all  $k\in \mathbb{N}$ and all $\mathbf{u} \in  \text{UC}_\mathbf{0}$. 
Since $\{ \mu^k \}_{k \in \mathbb{N}}$ converges to $\mu$, we have 
$\sum_{i\in N} \mu_i  \mathbf{u}_i \geq 0$ for all $\mathbf{u} \in  \text{UC}_\mathbf{0}$, 
that is, $\mu\in \mathcal{M}$. 

We prove that for any $\mathbf{u}\in \mathbb{R}^N$,  
\begin{equation}
\label{eq:equiv_tildepsi}
    \widetilde{\psi}(\mathbf{u})\geq 0 \iff \min_{\mu\in \mathcal{M}} \sum_{i\in N} \mu_i  \mathbf{u}_i \geq 0. 
\end{equation} 
To see this, let $\mathbf{u}\in \mathbb{R}^N$ with $\widetilde{\psi} (\mathbf{u})\geq 0$. Then,   we have  $\sum_{i\in N} \mu_i  \mathbf{u}_i \geq 0$ for all $\mu\in \mathcal{M}$, that is, $\min_{\mu\in \mathcal{M}} \sum_{i\in N} \mu_i  \mathbf{u}_i \geq 0$.
For the converse, assume that there exists $\mathbf{u}\in \mathbb{R}^N$ such that $\min_{\mu\in \mathcal{M}} \sum_{i\in N} \mu_i \mathbf{u}_i \geq 0$ but $\widetilde{\psi}(\mathbf{u}) <  0$ (i.e., $\mathbf{u} \notin \text{UC}_\mathbf{0}$). 
Then, by the separating hyperplane theorem, there exist $\mu'\in \Delta_N$ and $\kappa\in \mathbb{R}$ such that $\min_{\mathbf{u}'\in \text{UC}_\mathbf{0}} \sum_{i\in N}\mu'_i  \mathbf{u}_i \geq \kappa > \sum_{i\in N} \mu'_i  \mathbf{u}_i$. Since $\text{UC}_\mathbf{0}$ is a cone, we can set $\kappa = 0$. 
By the definition of $\mathcal{M}$, we have $\mu'\in \mathcal{M}$, which is a contradiction to $\min_{\mu\in \mathcal{M}} \sum_{i\in N} \mu_i \mathbf{u}_i \geq 0$.

We then prove that for all $\mathbf{u}\in \mathbb{R}^N$ with $\widetilde{\psi}(\mathbf{u}) = 0$,  $\min_{\mu\in \mathcal{M}} \sum_{i\in N} \mu_i  \mathbf{u}_i = 0$ holds.
By \eqref{eq:equiv_tildepsi}, it is sufficient to prove that $\min_{\mu\in \mathcal{M}} \sum_{i\in N} \mu_i  \mathbf{u}_i > 0$ does not hold. 
Suppose to the contrary that $\min_{\mu\in \mathcal{M}} \sum_{i\in N} \mu_i  \mathbf{u}_i > 0$. 
Let $\varepsilon \in \mathbb{R}_{++}$ with $0 < \varepsilon < \min_{\mu\in \mathcal{M}}  \sum_{i\in N} \mu_i  \mathbf{u}_i$. 
Then, we have  $\min_{\mu\in \mathcal{M}} \sum_{i\in N} \mu_i ( \mathbf{u}_i - \varepsilon)  > 0$. 
By the result of the last paragraph, we have $\widetilde{\psi}(\mathbf{u} - \varepsilon \mathbf{1}) \geq 0$. 
Since $\widetilde{\psi}$ is monotonic, we have $0 = \widetilde{\psi}(\mathbf{u}) > \widetilde{\psi}(\mathbf{u} - \varepsilon \mathbf{1}) \geq 0$, which is a contradiction. 

Finally, we prove that $\widetilde{\psi}$ can be written as $\widetilde{\psi} (\mathbf{u}) = \min_{\mu\in \mathcal{M}}  \sum_{i\in N} \mu_i \mathbf{u}_i$ for all $\mathbf{u} \in \mathbb{R}^N$. 
For each $\mathbf{u} \in \mathbb{R}^N$, since $\widetilde{\psi}$ is translation-invariant, we can take $\mathbf{u}^\ast \in \mathbb{R}^N$ such that $\widetilde{\psi} (\mathbf{u}^\ast) = 0$ and $\mathbf{u} = \mathbf{u}^\ast + \widetilde{\psi} (\mathbf{u}) \mathbf{1}$. 
Therefore, by the result in the last paragraph, 
\begin{align*}
    \widetilde{\psi} (\mathbf{u}) 
    &= \widetilde{\psi} \left(\mathbf{u}^\ast +   \widetilde{\psi}(\mathbf{u}) \mathbf{1}\right) 
    = \widetilde{\psi} (\mathbf{u}^\ast) + \widetilde{\psi} \left(\widetilde{\psi} (\mathbf{u}) \mathbf{1}\right) 
    = \left( \min_{\mu\in \mathcal{M}} \sum_{i\in N} \mu_i {\mathbf{u}}^\ast_i \right) + \widetilde{\psi} (\mathbf{u}) \\
    &
    = \min_{\mu\in \mathcal{M}} \sum_{i\in N} \mu_i \left(\mathbf{u}^\ast_i +\widetilde{\psi} (\mathbf{u}) \right)
    = \min_{\mu\in \mathcal{M}}  \sum_{i\in N} \mu_i  \mathbf{u}_i, 
\end{align*}
where the second equality follows from the translation-invariance of $\widetilde{\psi}$. \qed

\subsection{Independence of the axioms in Theorem~\ref{thm_relativefair}}
\label{app_ind}

\paragraph*{Dropping weak Pareto principle.}
Consider the ex-post aggregation rule $\mathbf{R}$ that follows the SEU preference associated with the average of individual beliefs and the average of 0--1 normalized utility functions; that is, for each $(X,R_N)\in\mathcal{D}$, $\mathbf{R}(X,R_N)$ is represented by the function $W_{(X,R_N)} : F_X \to\mathbb{R}$ defined by, for all $f\in F_X$,
$$W_{(X,R_N)}(f)=\int_{\Omega}  {\sum_{i \in N} u^{\ast}_i (f(\omega); X, R_N) \over |N|} d\bar{p}(\omega),$$
where $\bar{p}=\frac{1}{|N|} \sum_{i \in N} p_i$ (see \citet{brandl2021belief} for the corresponding rule in a different domain). 
Recall the impossibility result shown by \cite{mongin1995consistent}: Any non-dictatorial social preference does not satisfy the Paretian requirement if both the individuals and the social planner are SEU maximizers and the preference profile does not fulfill certain conditions.
Thus, the above rule violates \textit{weak Pareto principle} but satisfies all the other axioms. 

\paragraph*{Dropping event continuity.}
For $\mathbf{u}\in [0,1]^N$ and $i\in N$, let $\mathbf{u}_{(i)}$ be the $i$-th smallest element in $\mathbf{u}$, where ties are broken arbitrarily. 
Let $\geq_\text{lex}$ be the binary relation over $[0,1]^N$ such that for all $\mathbf{u},\mathbf{v}\in [0,1]^N$, 
\begin{align*}
    \mathbf{u} >_\text{lex} \mathbf{v} &\iff [ ~ \exists j \in  N  ~~\text{s.t.} ~~ \mathbf{u}_{(i)} =\mathbf{v}_{(i)} ~~\text{for all $i\in \{1,2,\ldots, j-1\}$ and} ~~ \mathbf{u}_{(j)} > \mathbf{v}_{(j)} ],
    \\
    \mathbf{u} =_\text{lex} \mathbf{v} &\iff [ ~ \mathbf{u}_{(i)} = \mathbf{v}_{(i)} ~~\text{for all $i\in N$} ~ ],  
\end{align*}
where $>_\text{lex}$ and $=_\text{lex}$ are the asymmetric and symmetric parts of $\geq_\text{lex}$, respectively. 
The \textit{relative leximin aggregation rule} is the aggregation rule $\mathbf{R}$ such that for all $(X,R_N) \in \mathcal{D}$ and all  $f, g \in F_X$, 
\begin{equation}
    f \, \mathbf{R}(X, R_N) \, g
    \iff 
    U^\ast (f;  X, R_N) 
    \geq_\text{lex}
    U^\ast (g; X, R_N).  
\end{equation}
This rule violates \textit{event continuity} but satisfies all the other axioms. 

\paragraph*{Dropping belief irrelevance.}
A belief-weighted relative utilitarian aggregation rule, where the weights over the individuals change depending on the belief profile, violates \textit{belief irrelevance} in general but satisfies all the other axioms (\cite{sprumont2019relative}).

\paragraph*{Dropping independence of redundant outcomes.}
Consider the following aggregation rule $\mathbf{R}$: If the number of elements in $X$ is odd, 
the aggregation rule coincides with some relative utilitarian aggregation rule (Definition \ref{def:relativeutilitarian}); otherwise, it coincides with the relative maximin aggregation rule (Definition \ref{def:relativemaximin}).
By construction, for any $X\in\mathcal{X}$ and $X'=X \cup \{x\}$ such that $x \in \mathbb{X} \setminus X$, the evaluation on $F_X$ by the aggregation rule $\mathbf{R}$ with $X$ is not always consistent with the evaluation on $F_X (\subset F_{X'})$ by $\mathbf{R}$ with $X'$.
Thus, this aggregation rule violates \textit{independence of redundant outcomes}  but satisfies all the other axioms. 

\paragraph*{Dropping weak preference for mixing.}
Consider an aggregation rule $\mathbf{R}$ such that larger weights are assigned to those who enjoy higher normalized utility; that is, for each $(X,R_N)\in\mathcal{D}$,  $\mathbf{R}(X,R_N)$ is represented by the function $W_{(X,R_N)}: F_X \to\mathbb{R}$ defined by, for all $f\in F_X$,
$$W_{(X,R_N)}(f)=\max_{\mu \in \mathcal{M}} \sum_{i\in N} \mu_i U_i^\ast (f; X, R_N),$$
where $\mathcal{M}$ is a nonempty closed subset of $\Delta_N$, as in the relative fair aggregation rules. 
If $\mathcal{M}$ is not a singleton, this violates \textit{weak preference for mixing} but satisfies all the other axioms.

\paragraph*{Dropping restricted certainty independence.}
The Nashian aggregation rule, which evaluates each act based on the product of all individuals' utility levels, violates \textit{restricted certainty independence} but satisfies all the other axioms.\footnote{\cite{sprumont2018belief} characterized the belief-weighted Nashian aggregation rules where the weight vector depends on the belief profile.}

\subsection{Proof of Lemma~\ref{lem:addAnonym}}

Let $\mathbf{R}$ be an aggregation rule that satisfies 
\textit{weak Pareto principle}, \textit{event continuity}, \textit{belief irrelevance}, \textit{independence of redundant outcomes}, and \textit{anonymity}. 
Let $\psi :[0,1]^N \rightarrow \mathbb{R}$ be the monotonic continuous function in Lemma~\ref{lem:basic}. 
Define the binary relation $\succsim$ over $[0,1]^N$ as in the proof of Lemma~\ref{lem:quasiconcav}. 

Let $\mathbf{u}, \mathbf{v} \in [0,1]^N$ be such that for some $i,j\in N$, $\mathbf{u}_i =  \mathbf{v}_j$,  $\mathbf{u}_j=  \mathbf{v}_i$, and $\mathbf{u}_k =  \mathbf{v}_k$ for all $k\in N\backslash \{ i, j \}$. 
It suffices to prove that $\mathbf{u} \sim  \mathbf{v}$. Let $\pi \in \Pi$ be such that $\pi (i) = j$,  $\pi (j) = i$, and $\pi (k) = k$ for all $k\in N\backslash \{ i, j \}$. 
Suppose to the contrary that $\mathbf{u} \not\sim  \mathbf{v}$. 
We assume $\mathbf{u} \succ  \mathbf{v}$ without loss of generality. 
Take $( X, R_N ) \in \mathcal{D}$ and $x,y \in X$ such that $u^\ast (x; X, R_N) =  \mathbf{u}$, and $u^\ast (y ; X, R_N) =\mathbf{v}$. By the definition of $\succsim$, $x \,\mathbf{P}(X,R_N)\, y$. By \textit{anonymity}, $x \,\mathbf{P}(X, R_N^{\pi})\, y$. 
Note that $u^\ast (x; X, R_N^{\pi}) =  \mathbf{v}$ and $u^\ast (y ; X, R_N^{\pi}) =\mathbf{u}$. By the definition of $\succsim$, $\mathbf{v} \succ \mathbf{u}$, which is a contradiction to $\mathbf{u} \succ  \mathbf{v}$. 
\qed

\subsection{Proof of Theorem~\ref{thm:rel-util}}

Let $\mathbf{R}$ be an aggregation rule that satisfies \textit{weak Pareto principle}, \textit{event continuity}, \textit{belief irrelevance}, \textit{independence of redundant outcomes}, and \textit{certainty independence}. 
Let $\psi :[0,1]^N \rightarrow \mathbb{R}$ be the monotonic continuous function in Lemma~\ref{lem:basic}. 
Without loss of generality, we assume that for all $c \in [0,1]$, $\psi (c\mathbf{1}) = c$. 

Take $\mathbf{u}, \mathbf{v}, \mathbf{w} \in [0,1]^N$ and $\alpha \in (0,1)$ arbitrarily. 
We prove that $\psi (\mathbf{u} )\geq \psi(\mathbf{v})$ if and only if $\psi(\alpha \mathbf{u} + (1 - \alpha) \mathbf{w} ) \geq \psi ( \alpha \mathbf{v} + (1 - \alpha) \mathbf{w})$. 
Let $(X, R_N) \in \mathcal{D}$, $f, g \in F_X$, and $x\in X$ be such that $U^\ast (f;  X, R_N) = \mathbf{u}$, $U^\ast (g;   X, R_N) = \mathbf{v}$, and $u^\ast (x;   X, R_N) = \mathbf{w}$. 
By the definition of $\psi$, $\psi (\mathbf{u} )\geq \psi(\mathbf{v})$ is equivalent to $f \, \mathbf{R} (X, R_N) \, g$. 
By \textit{certainty independence}, this is equivalent to $f_{\alpha} x \, \mathbf{R} (X, R_N) \, g_{\alpha} x$.
Since  $R_i$ is an SEU preference for each $i\in N$, we have $U^\ast (f_{\alpha} x;   X, R_N)  = \alpha \mathbf{u} + (1 - \alpha) \mathbf{w}$ and $U^\ast (g_{\alpha} x;  X, R_N) = \alpha \mathbf{v} + (1 - \alpha) \mathbf{w}$. 
Therefore, $f \, \mathbf{R} (X, R_N) \, g$ is equivalent to  $\psi (\alpha \mathbf{u} + (1 - \alpha) \mathbf{w}) \geq \psi( \alpha \mathbf{v} + (1 - \alpha) \mathbf{w})$. 

Then, by applying Theorem~8 of \citet{herstein1953axiomatic}, there exist $\mu\in \mathbb{R}^N$ and an increasing function $\zeta: \mathbb{R}\to\mathbb{R}$ such that for all $\mathbf{u}\in [0,1]^N$, $\psi (\mathbf{u}) = \zeta \qty(\sum_{i\in N} \mu_i  \mathbf{u}_i)$.%
\footnote{More precisely, \citet{herstein1953axiomatic} imposed axioms not on a function but on a binary relation. By translating the functional properties that we have derived into their counterparts for binary relations, we can apply Herstein and Milnor's result.} 
By the monotonicity of $\psi$, we can see that $\mu\in \mathbb{R}^N_+$.
Thus, without loss of generality, we can set $\mu\in \Delta_N$. 
For each $c \in [0,1]$, it follows from $\psi (c\mathbf{1}) = c$ that $\zeta(c) = \zeta \qty(\sum_{i\in N} \mu_i c) = \psi (c\mathbf{1}) = c$.
Thus, $\psi (\mathbf{u}) = \sum_{i\in N} \mu_i  \mathbf{u}_i$ for all $\mathbf{u}\in [0,1]^N$.
\qed

\subsection{Proof of Theorem~\ref{thm_relmaxmin_mix}}

Let $\mathbf{R}$ be an aggregation rule that satisfies 
\textit{weak Pareto principle}, \textit{event continuity}, \textit{belief irrelevance}, \textit{independence of redundant outcomes}, \textit{anonymity}, and \textit{strong preference for mixing}. 
Let $\psi :[0,1]^N \rightarrow \mathbb{R}$ be the monotonic continuous function in Lemma~\ref{lem:basic}. 
Without loss of generality, we assume that for all $c \in [0,1]$, $\psi (c\mathbf{1}) = c$.
By \textit{anonymity}, $\psi$ is symmetric, that is, for all $\mathbf{u} \in [0,1]^N$ and all $\pi \in \Pi$,  $\psi (\mathbf{u}) = \psi (\mathbf{u}^\pi)$. 
For $\mathbf{u}, \mathbf{v} \in [0,1]^N$, let $\mathbf{u} \land \mathbf{v} = (\min \{ \mathbf{u}_1, \mathbf{v}_1 \}, \ldots, \min \{ \mathbf{u}_n, \mathbf{v}_n \} )$.

\begin{claim}
\label{claim:comp}
    For all $\mathbf{u}, \mathbf{v}, \mathbf{w} \in [0,1]^N$, if $\mathbf{w}\geq \mathbf{u} \land \mathbf{v}$, then $\psi (\mathbf{w}) \geq \psi (\mathbf{u}) $ or $\psi (\mathbf{w}) \geq \psi (\mathbf{v})$. 
\end{claim}

\begin{proof}
    \renewcommand{\qedsymbol}{$||$}
    Let  $\mathbf{u}, \mathbf{v}, \mathbf{w} \in [0,1]^N$ be such that $\mathbf{u} \land \mathbf{v} \leq \mathbf{w}$. 
    By the monotonicity of $\psi$, it suffices to consider the case with $\mathbf{w}_i \leq \max \{ \mathbf{u}_i, \mathbf{v}_i \} $ for all $i\in N$.
    Take $(X,R_N) \in \mathcal{D}$ and $x,y \in X$ such that (i) $U^\ast(x;X,R_N)=\mathbf{u}$ and  $U^\ast(y;X,R_N)=\mathbf{v}$ and (ii) there exists $E \subset \Omega$ such that for each $i\in N$, 
    \begin{equation*}
        p_i (E) = 
        \begin{cases}
            \displaystyle{ \mathbf{w}_i - \mathbf{v}_i \over \mathbf{u}_i - \mathbf{v}_i}  & \text{if $\mathbf{u}_i > \mathbf{v}_i$} \\ 
            1 & \text{if $\mathbf{u}_i = \mathbf{v}_i$} \\
            \displaystyle{ \mathbf{v}_i - \mathbf{w}_i \over \mathbf{v}_i - \mathbf{u}_i } & \text{if $\mathbf{u}_i <  \mathbf{v}_i$}
        \end{cases}.
    \end{equation*}
    Then, for $i\in N$ with $\mathbf{u}_i > \mathbf{v}_i$, 
     \begin{equation*}
        U_i^\ast(xEy;X,R_N)=  \qty({\mathbf{w}_i - \mathbf{v}_i \over \mathbf{u}_i - \mathbf{v}_i }) \mathbf{u}_i  + \qty( 1 -  {\mathbf{w}_i - \mathbf{v}_i \over \mathbf{u}_i - \mathbf{v}_i })\mathbf{v}_i = \mathbf{w}_i. 
    \end{equation*}
    For $i\in N$ with $\mathbf{u}_i = \mathbf{v}_i$, $U_i^\ast(xEy;X,R_N)= \mathbf{u}_i = \mathbf{v}_i =  \mathbf{w}_i$. 
    Moreover, for $i\in N$ with $\mathbf{u}_i < \mathbf{v}_i$, 
    \begin{equation*}
        U_i^\ast(xEy;X,R_N)=  \qty({\mathbf{v}_i - \mathbf{w}_i \over \mathbf{v}_i - \mathbf{u}_i }) \mathbf{u}_i  + \qty( 1 -  {\mathbf{v}_i - \mathbf{w}_i \over \mathbf{v}_i - \mathbf{u}_i })\mathbf{v}_i = \mathbf{w}_i. 
    \end{equation*}
    Therefore, $U^\ast(xEy;X,R_N) = \mathbf{w}$. By \textit{strong preference for mixing}, $xEy\, \mathbf{R}(X, R_N ) \, x$ or $xEy \,  \mathbf{R} (X, R_N ) \, y$. 
    By the definition of $\psi$, $\psi(\mathbf{w}) \geq \psi(\mathbf{u})$ or $\psi( \mathbf{w}) \geq \psi( \mathbf{v}) $.
\end{proof}

By the symmetry of $\psi$,
it is sufficient to prove that $\psi (\mathbf{u}) = \mathbf{u}_1$ for all $\mathbf{u} \in [0,1]^N$ such that $\mathbf{u}_1 \leq \mathbf{u}_2 \leq \cdots \leq \mathbf{u}_n$. 
The monotonicity of $\psi$ implies $\psi (\mathbf{u}) \geq \mathbf{u}_1$. 

By the symmetry of $\psi$, we have $ \psi (\mathbf{u}_1,  \mathbf{u}_2, \mathbf{u}_3, \ldots, \mathbf{u}_n) = \psi (\mathbf{u}_2,  \mathbf{u}_1, \mathbf{u}_3, \ldots, \mathbf{u}_n)$. 
Since $ (\mathbf{u}_1,  \mathbf{u}_2,\mathbf{u}_3, \ldots, \mathbf{u}_n) \land (\mathbf{u}_2,  \mathbf{u}_1,\mathbf{u}_3, \ldots, \mathbf{u}_n) = (\mathbf{u}_1,  \mathbf{u}_1,\mathbf{u}_3, \ldots, \mathbf{u}_n)$, Claim~\ref{claim:comp} implies 
\begin{equation*}
    \psi (\mathbf{u}_1,  \mathbf{u}_1,\mathbf{u}_3, \ldots, \mathbf{u}_n) \geq \psi (\mathbf{u}_1,  \mathbf{u}_2, \mathbf{u}_3,  \ldots, \mathbf{u}_n).
\end{equation*} 
By this inequality and the symmetry of $\psi$, $\psi (\mathbf{u}_1,  \mathbf{u}_1,\mathbf{u}_3, \ldots, \mathbf{u}_n) \geq \psi (\mathbf{u}_3,  \mathbf{u}_2, \mathbf{u}_1, \mathbf{u}_4, \ldots, \mathbf{u}_n)$ holds.  
Since $ (\mathbf{u}_1,  \mathbf{u}_1,\mathbf{u}_3, \ldots, \mathbf{u}_n) \land (\mathbf{u}_3,  \mathbf{u}_2, \mathbf{u}_1, \mathbf{u}_4,  \ldots, \mathbf{u}_n) = (\mathbf{u}_1,  \mathbf{u}_1,\mathbf{u}_1, \mathbf{u}_4, \ldots, \mathbf{u}_n)$, Claim~\ref{claim:comp} and the symmetry of $\psi$ imply
\begin{equation*}
    \psi (\mathbf{u}_1,  \mathbf{u}_1,\mathbf{u}_1,  \mathbf{u}_4, \ldots, \mathbf{u}_n) \geq \psi (\mathbf{u}_3,  \mathbf{u}_2, \mathbf{u}_1, \mathbf{u}_4,  \ldots, \mathbf{u}_n)= \psi (\mathbf{u}_1,  \mathbf{u}_2, \mathbf{u}_3,  \ldots, \mathbf{u}_n). 
\end{equation*} 
By repeating this argument, we have $\psi (\mathbf{u}_1,  \mathbf{u}_1,\mathbf{u}_1, \ldots, \mathbf{u}_1) \geq \psi (\mathbf{u}_1,  \mathbf{u}_2, \mathbf{u}_3,  \ldots, \mathbf{u}_n)$, that is, $\mathbf{u}_1 \geq \psi (\mathbf{u}) $. 
\qed

\subsection{Proof of Theorem~\ref{thm_ambavo_maxmin}}

Let $\mathbf{R}$ be an aggregation rule that satisfies \textit{weak Pareto principle}, \textit{event continuity}, \textit{belief irrelevance}, \textit{independence of redundant outcomes}, and \textit{social ambiguity avoidance}. 
Let $\psi :[0,1]^N \rightarrow \mathbb{R}$ be the monotonic continuous function in Lemma~\ref{lem:basic}. 
Without loss of generality, we assume that for all $c \in [0,1]$, $\psi (c\mathbf{1}) = c$.
Let $\mathbf{u} \in [0,1]^N$.
By the continuity and monotonicity of $\psi$, there exists $c_\mathbf{u} \in [0,1]$ such that $\psi (\mathbf{u}) = \psi (c_\mathbf{u} \mathbf{1})$. 

First, we verify $\min_{i\in N}  \mathbf{u}_i \geq c_\mathbf{u}$. Suppose to the contrary that for some $i^\ast \in N$, $ \mathbf{u}_{i^\ast} < c_\mathbf{u}$. 
Then there exist $(X, R_N) \in \mathcal{D}$, $x\in X$, and $f\in F_X$ such that $u_i^\ast (\cdot;  X, R_N) = u_j^\ast (\cdot;  X, R_N)$ for each $i, j \in N$,  $u^\ast  (x; X, R_N) = c_\mathbf{u} \mathbf{1}$,  and $U^\ast  (f;  X, R_N) = \mathbf{u}$.  
By $ \mathbf{u}_{i^\ast} < c_\mathbf{u}$, we have $x P_{i^\ast} f$. 
By \textit{social ambiguity avoidance}, $x \,\mathbf{P} (X, R_N) \,f$, which implies $\psi (c_\mathbf{u} \mathbf{1}) > \psi (\mathbf{u})$. 
This is a contradiction to the definition of $c_\mathbf{u}$. 

Then, we prove $\min_{i\in N}  \mathbf{u}_i = c_\mathbf{u}$, that is, $\psi (\mathbf{u}) = \min_{i\in N}  \mathbf{u}_i$.
Suppose to the contrary that $\min_{i\in N}  \mathbf{u}_i > c_\mathbf{u} = \psi (c_\mathbf{u} \mathbf{1})$. 
By the definition of $c_\mathbf{u}$, we have $\min_{i\in N}  \mathbf{u}_i >  \psi (c_\mathbf{u} \mathbf{1}) = \psi ( \mathbf{u})$.
Since $\psi$ is monotonic, $\psi (\mathbf{u} )  \geq   \psi \qty( (\min_{i\in N}  \mathbf{u}_i) \mathbf{1} ) = \min_{i\in N}  \mathbf{u}_i$, which is a contradiction. \qed

\section{Generalization}
\label{app_general}

Theorems~\ref{thm_relativefair} and \ref{thm:rel-util} characterize classes of aggregation rules with the axioms corresponding to \citeauthor{gilboa1989maxmin}'s (\citeyear{gilboa1989maxmin}) certainty independence.
Here, we study a general class of aggregation rules, focusing on the independence property.
In the literature on decision-making under uncertainty, \citet{maccheroni2006ambiguity} proposed a weaker independence axiom, which they call ``weak certainty independence,'' to generalize the maxmin EU model.
The following is the counterpart of this axiom in our setup. 

\vspace{1mm}

\begin{description}
\item[\bf Weak Restricted Certainty Independence.] For all $(X, R_N) \in \mathcal{D}$ such that $u_i^\ast (\cdot;  X, R_N) = u_j^\ast (\cdot;  X, R_N)$ for each $i,j\in N$, all $f, g \in F_X$, all $x, y\in X$, and all $\alpha \in (0,1)$, $f_{\alpha} x \, \mathbf{R}(X, R_N ) \, g_{\alpha} x$ if and only if $f_{\alpha} y \, \mathbf{R}(X, R_N ) \, g_{\alpha} y$. 
\end{description}

\vspace{1mm}

Note that, like \textit{restricted certainty independence}, this axiom focuses on problems where all individuals share a common 0--1 normalized value function. 

By replacing \textit{restricted certainty independence} in Theorem~\ref{thm_relativefair} with the above axiom, we obtain a general class of aggregation rules. 
We say that a function $\varphi: \Delta_N \rightarrow \mathbb{R}\cup \{+\infty\}$ is \textit{grounded} if $\inf_{\mu \in \Delta_N}\varphi(\mu) = 0$.

\begin{theorem}
\label{thm_gene}
 An aggregation rule $\mathbf{R}$ satisfies \textnormal{weak Pareto principle}, \textnormal{event continuity}, \textnormal{belief irrelevance}, \textnormal{independence of redundant outcomes}, \textnormal{weak preference for mixing},  and \textnormal{weak restricted certainty independence} if and only if there exists a grounded, convex, lower semicontinuous function $\varphi :  \Delta_N \rightarrow \mathbb{R}_+\cup \{+\infty\}$ such that for each $(X, R_N) \in \mathcal{D}$, $\mathbf{R} (X, R_N)$ is represented by the function $W_{(X, R_N)} :F_X \to \mathbb{R}$ defined by, for all $f\in F_X$,
\begin{equation}
\label{eq_variational}
    W_{(X, R_N)} (f) = \min_{\mu\in   \Delta_N} \left\{\sum_{i\in N} \mu_i U_i^\ast (f; X, R_N) + \varphi(\mu) \right\}.  
\end{equation}
\end{theorem}

The function $\varphi$ in \eqref{eq_variational} represents how (un)reasonable the social planner thinks it is to take each weight vector into account: The higher $\varphi(\mu)$ is, the less reasonable the social planner thinks the weight vector $\mu$ to be. 
Under the aggregation rule \eqref{eq_variational}, the planner evaluates each act as follows: First, the planner normalizes each individual's utility function; second, for each weight vector $\mu\in \Delta_N$, the planner computes the weighted sum of the normalized utility levels and then adds the value $\varphi(\mu)$; finally, to determine the weights to be used for evaluation, the planner chooses the minimum among the values computed in the second step. As in the final step of the relative fair aggregation rules, the final part corresponds to the egalitarian attitude of the planner. 

\subsection{Proof of Theorem~\ref{thm_gene}}

Let $\mathbf{R}$ be an aggregation rule that satisfies 
\textit{weak Pareto principle}, \textit{event continuity}, \textit{belief irrelevance}, \textit{independence of redundant outcomes}, \textit{weak preference for mixing}, and \textit{weak restricted certainty independence}. 
Let $\psi :[0,1]^N \rightarrow \mathbb{R}$ be the monotonic, continuous, and quasiconcave function in Lemma~\ref{lem:quasiconcav}. 
Without loss of generality, we assume that for all $c \in [0,1]$, $\psi (c\mathbf{1}) = c$. 

\begin{claim}
\label{claim_niv}
    The function $\psi$ is translation-invariant and concave. 
\end{claim}

\begin{proof}
    \renewcommand{\qedsymbol}{$||$}
    Let $\mathbf{u} \in [0,1]^N$, $c\in [0,1]$, and $\alpha \in (0,1)$. 
    Since $\psi$ is monotonic and continuous, there exists $c^\ast \in [0,1]$ such that 
\begin{equation}
\label{eq:quasi_equi}
    \psi (\alpha \mathbf{u} + (1 -\alpha ) c\mathbf{1}) = \psi (\alpha c^\ast \mathbf{1} + (1 -\alpha ) c\mathbf{1}) . 
\end{equation}
Take $( X, R_N ) \in \mathcal{D}$ and $f\in F_X$ such that (i)  $u_i^\ast (\cdot;  X, R_N) = u_j^\ast (\cdot;  X, R_N)$ for each $i,j\in N$, 
(ii) for some $x, y \in X$, $c \mathbf{1} = U^\ast(x; X, R_N)$ and $c^\ast \mathbf{1} = U^\ast(y; X, R_N)$, (iii) there exists  $x_\ast \in X$ with $z R_i x_\ast$ for all $i\in N$ and all $z \in X$, and (iv) $U^\ast(f ; X, R_N) = \mathbf{u}$.\footnote{We can construct such an act $f$ in a way similar to the argument in the proof of Lemma~\ref{lem:4addCI}.}
Note that $U^\ast(f_{\alpha} x ; X, R_N) = \alpha \mathbf{u} + (1- \alpha ) c\mathbf{1}$ and $U^\ast(y_{\alpha} x ; X, R_N) = \alpha c^\ast \mathbf{1} + (1- \alpha ) c\mathbf{1}$. 
By \eqref{eq:quasi_equi} and the definition of $\psi$, 
$f_{\alpha} x \, \mathbf{I}(X, R_N ) \, y_{\alpha} x$. 
By \textit{weak restricted certainty independence}, $f_{\alpha} x_\ast \, \mathbf{I}(X, R_N ) \, y_{\alpha} x_\ast$. 
Since $U^\ast(f_{\alpha} x_\ast ; X, R_N) = \alpha \mathbf{u}$ and $U^\ast(y_{\alpha} x_{\ast} ; X, R_N) = \alpha c^\ast \mathbf{1}$, $\psi(\alpha \mathbf{u} ) = \psi ( \alpha c^\ast \mathbf{1} ) =  \alpha c^\ast$. 
Therefore,
\begin{equation}
\label{eq:quasi-tansinv}
    \psi (\alpha \mathbf{u} + (1 -\alpha ) c\mathbf{1})
    = \psi (\alpha c^\ast \mathbf{1} + (1 -\alpha ) c\mathbf{1}) 
    = \alpha c^\ast +(1- \alpha)c =  \psi(\alpha \mathbf{u}) +(1- \alpha)c, 
\end{equation}
where the second equality follows from the assumption that for all $c \in [0,1]$, $\psi (c\mathbf{1}) = c$ and the fact that $\alpha c^\ast +(1- \alpha)c \in [0,1]$. 
By Theorem~4 of \citet{cerreia2014niveloids}, $\psi$ is translation-invariant.

Since $\psi$ is quasiconcave, for all $\mathbf{v}, \mathbf{w}$ with $\psi (\mathbf{v}) = \psi (\mathbf{w})$ and all $\beta \in (0,1)$, $\psi (\beta \mathbf{v} + (1-\beta )\mathbf{w}) \geq \psi (\mathbf{v})$. 
By Theorem~4 of \citet{cerreia2014niveloids}, $\psi$ is concave. 
\end{proof}

Let $\mathbf{u} \in (0,1)^N$. 
Since $-\psi$ is convex (cf.\ Claim~\ref{claim_niv}),  Theorem~7.12 of \citet{AB2006Math} implies that  $-\psi$ is subdifferentiable at $\mathbf{u}$: 
That is, there exists $\mu^\ast\in \mathbb{R}^N$ such that for all $\mathbf{v}\in (0,1)^N$, 
\begin{equation}
\label{eq:subdif}
    \psi (\mathbf{v}) - \sum_{i\in N} \mu^\ast_i \mathbf{v}_i  \leq \psi (\mathbf{u}) - \sum_{i\in N} \mu^\ast_i \mathbf{u}_i. 
\end{equation}

\begin{claim}
\label{claim:prob}
     $\mu^\ast \in\Delta_N$ holds.
\end{claim}

\begin{proof}
\renewcommand{\qedsymbol}{$||$}
First, we prove that $\mu^\ast\geq 0$. 
Fix $k \in N$ arbitrarily. Let $\mathbf{v} \in (0,1)^N$ be such that $\mathbf{v}_k > \mathbf{u}_k$ and $\mathbf{v}_j = \mathbf{u}_j$ for all $j\in N\backslash \{k \}$. 
Since $\psi$ is monotonic and continuous, \eqref{eq:subdif} implies that $0\leq \psi (\mathbf{v}) - \psi (\mathbf{u}) \leq \sum_{i\in N} \mu^\ast_i (\mathbf{v}_i - \mathbf{u}_i) = (\mathbf{v}_k - \mathbf{u}_k) \mu^\ast_k$.
By $\mathbf{v}_k > \mathbf{u}_k$, we have $\mu^\ast_k \geq 0$. 

Next, we show that $\sum_{i\in N} \mu^\ast_i = 1$. Suppose to the contrary that $\sum_{i\in N} \mu^\ast_i \neq 1$. 
If $\sum_{i\in N} \mu^\ast_i < 1$, then let $c> 0$ be such that  $ \mathbf{u} + c\mathbf{1} \in (0,1)^N$. 
By Claim~\ref{claim_niv}, 
\begin{align*}
    \sum_{i\in N} \mu^\ast_i (\mathbf{u}_i  + c) - \psi(\mathbf{u} + c\mathbf{1}) 
    &= \bigg( \sum_{i\in N} \mu^\ast_i - 1 \bigg)  c +  \sum_{i\in N} \mu^\ast_i \mathbf{u}_i  - \psi(\mathbf{u}) < \sum_{i\in N} \mu^\ast_i \mathbf{u}_i  - \psi(\mathbf{u}), 
\end{align*} 
which contradicts \eqref{eq:subdif}. 
If $\sum_{i\in N} \mu^\ast_i > 1$, then let $c' <0$ be such that $\mathbf{u} + c' \mathbf{1} \in (0,1)^N$. 
Similarly, by Claim~\ref{claim_niv}, 
we have $\sum_{i\in N} \mu^\ast_i (\mathbf{u}_i + c') - \psi(\mathbf{u} + c' \mathbf{1}) < \sum_{i\in N} \mu^\ast_i \mathbf{u}_i  - \psi(\mathbf{u})$,
which contradicts \eqref{eq:subdif}.
\end{proof}
    
Let $\varphi : \Delta_N \rightarrow \mathbb{R}\cup\{ +\infty \}$ be the function such that for all $\mu\in \Delta_N$, 
$\varphi (\mu) = \sup_{\mathbf{u} \in (0,1)^N} \big\{ \psi(\mathbf{u}) -  \sum_{i\in N} \mu_i \mathbf{u}_i \big\}$.
By \eqref{eq:subdif} and Claim~\ref{claim:prob}, for all $\mathbf{u} \in (0,1)^N$, there exists $\mu^\ast\in\Delta_N$ such that $\psi (\mathbf{u}) = \sum_{i\in N } \mu^\ast_i \mathbf{u}_i + \varphi (\mu^\ast)$.  
Also, by the definition of $\varphi$, for all $\mu\in \Delta_N$ and all $\mathbf{u} \in (0,1)^N$, 
$ \varphi (\mu) \geq   \psi(\mathbf{u}) -  \sum_{i\in N} \mu_i \mathbf{u}_i $, that is, $\psi (\mathbf{u}) \leq  \sum_{i\in N} \mu_i \mathbf{u}_i + \varphi (\mu) $.
Thus, for all $\mathbf{u} \in (0,1)^N$, 
\begin{equation}
\label{eq_nivvar}
        \psi (\mathbf{u}) = \min_{\mu\in \Delta_N} \left\{ \sum_{i\in N} \mu_i \mathbf{u}_i + \varphi (\mu) \right\}. 
\end{equation}
By the construction, $\min_{\mu\in \Delta_N} \varphi (\mu) = 0$.  
By Lemma~5.40~(3) of \citet{AB2006Math}, the pointwise supremum of a family of linear functions is convex.
Therefore, $\varphi$ is a convex function. 
By Lemma~2.41 of \citet{AB2006Math}, the pointwise supremum of a family of lower semicontinuous functions is lower semicontinuous, which implies that $\varphi$ is a lower semicontinuous function. 

Note that by Theorem~4 of \citet{cerreia2014niveloids}, $\psi$ is 1-Lipschitz continuous with respect to the sup norm (i.e., $|\psi (\mathbf{u}) - \psi (\mathbf{v})| \leq \sup_{i\in N} |\mathbf{u}_i - \mathbf{v}_i|$ for all $\mathbf{u}, \mathbf{v} \in [0,1]^N$).
Thus, \eqref{eq_nivvar} holds for every point in $[0,1]^N$. 
\qed

\end{appendix}

\bibliographystyle{te}
\bibliography{ref}

@Book{AB2006Math,
  author={Charalambos D. Aliprantis and Kim C. Border},
  title={{Infinite Dimensional Analysis}},
  publisher={Berlin: Springer},
  year={2006},
  month={September},
  volume={},
  number={},
  series={},
  edition={},
  doi={},
  isbn={ARRAY(0x6344c4f0)},
  url={}
}

@article{alon2014purely,
  title={Purely subjective maxmin expected utility},
  author={Alon, Shiri and Schmeidler, David},
  journal={Journal of Economic Theory},
  volume={152},
  pages={382--412},
  year={2014},
  publisher={Elsevier}
}

@article{anscombe1963definition,
  title={A definition of subjective probability},
  author={Anscombe, Francis J. and Aumann, Robert J.},
  journal={Annals of Mathematical Statistics},
  volume={34},
  number={1},
  pages={199--205},
  year={1963}
}

@article{alon2016utilitarian,
  title={Utilitarian preferences with multiple priors},
  author={Alon, Shiri and Gayer, Gabi},
  journal={Econometrica},
  volume={84},
  number={3},
  pages={1181--1201},
  year={2016},
  publisher={Wiley Online Library}
}

@article{armstrong1981,
  title={Liapounoff's theorem for nonatomic, finitely-additive, bounded, finite-dimensional, vector-valued measures},
  author={Armstrong, T. E. and Prikry, Karel},
  journal={Transactions of the American Mathematical Society},
  volume={266},
  number={2},
  pages={499--514},
  year={1981},
  publisher={}
}

@book{arrow2012social,
  title={Social Choice and Individual Values},
  author={Arrow, Kenneth J.},
  year={1951},
  publisher={New York: Wiley}
}

@article{baris2018timing,
  title={Timing effect in bargaining and ex ante efficiency of the relative utilitarian solution},
  author={Baris, Omer F.},
  journal={Theory and Decision},
  volume={84},
  number = {4},
  pages={547--556},
  year={2018},
  publisher={Springer}
}

@article{ben1997measurement,
  title={On the measurement of inequality under uncertainty},
  author={Ben-Porath, Elchanan and Gilboa, Itzhak and Schmeidler, David},
  journal={Journal of Economic Theory},
  volume={75},
  number={1},
  pages={194--204},
  year={1997},
  publisher={Elsevier}
}

@article{brandl2021belief,
  title={Belief-averaging and relative utilitarianism},
  author={Brandl, Florian},
  journal={Journal of Economic Theory},
  volume={198},
  pages={105368},
  year={2021},
  publisher={Elsevier}
}

@unpublished{borgers2017revealed,
  title={Revealed relative utilitarianism},
  author={B{\"o}rgers, Tilman and Choo, Yan Min},
  year={2017},
 institution={University of Michigan},
  publisher={CESifo Working Paper Series},
  note = {working paper}
}

@article{borie2023maxmin,
  title={Maxmin expected utility in {S}avage's framework},
  author={Borie, Dino},
  journal={Journal of Economic Theory},
  volume={210},
  pages={105665},
  year={2023},
  publisher={Elsevier}
}

@article{casadesus2000maxmin,
  title={Maxmin expected utility over {S}avage acts with a set of priors},
  author={Casadesus-Masanell, Ramon and Klibanoff, Peter and Ozdenoren, Emre},
  journal={Journal of Economic Theory},
  volume={92},
  number={1},
  pages={35--65},
  year={2000},
  publisher={Elsevier}
}

@article{cerreia2014niveloids,
  title={Niveloids and their extensions: Risk measures on small domains},
  author={Cerreia-Vioglio, Simone and Maccheroni, Fabio and Marinacci, Massimo and Rustichini, Aldo},
  journal={Journal of Mathematical Analysis and Applications},
  volume={413},
  number={1},
  pages={343--360},
  year={2014},
  publisher={Elsevier}
}

@article{chambershayashi2006preference,
  title={Preference aggregation under uncertainty: {S}avage vs. {P}areto},
  author={Chambers, Christopher P. and Hayashi, Takashi},
  journal={Games and Economic Behavior},
  volume={54},
  number={2},
  pages={430--440},
  year={2006},
  publisher={Elsevier}
}

@article{cres2011aggregation,
  title={Aggregation of multiple prior opinions},
  author={Cr{\`e}s, Herv{\'e} and Gilboa, Itzhak and Vieille, Nicolas},
  journal={Journal of Economic Theory},
  volume={146},
  number={6},
  pages={2563--2582},
  year={2011},
  publisher={Elsevier}
}

@article{dAspremont1977equity,
	author = {d'Aspremont, Claude and Gevers, Louis},
	journal = {Review of Economic Studies},
	number = {2},
	pages = {199--209},
	title = {Equity and the informational basis of collective choice},
	volume = {44},
	year = {1977}}

@article{danan2016robust,
  title={Robust social decisions},
  author={Danan, Eric and Gajdos, Thibault and Hill, Brian and Tallon, Jean-Marc},
  journal={American Economic Review},
  volume={106},
  number={9},
  pages={2407--2425},
  year={2016},
  publisher={American Economic Association 2014 Broadway, Suite 305, Nashville, TN 37203}
}

@article{diamond,
  title={Cardinal welfare, individualistic ethics, and interpersonal comparison of utility: Comment},
  author={Diamond, Peter A.},
  journal={Journal of Political Economy},
  volume={75},
  number={5},
  pages={765--766},
  year={1967},
  publisher={The University of Chicago Press}
}

@article{dhillon1999relative,
  title={Relative utilitarianism},
  author={Dhillon, Amrita and Mertens, Jean-Fran{\c{c}}ois},
  journal={Econometrica},
  volume={67},
  number={3},
  pages={471--498},
  year={1999},
  publisher={Wiley Online Library}
}

@article{fleurbaey2021fair,
  title={Fair utilitarianism},
  author={Fleurbaey, Marc and Zuber, St{\'e}phane},
  journal={American Economic Journal: Microeconomics},
  volume={13},
  number={2},
  pages={370--401},
  year={2021},
  publisher={American Economic Association 2014 Broadway, Suite 305, Nashville, TN 37203-2425}
}

@article{gilboa1989maxmin,
  title={Maxmin expected utility with non-unique prior},
  author={Gilboa, Itzhak and Schmeidler, David},
  journal={Journal of Mathematical Economics},
  volume={18},
  number={2},
  pages={141--153},
  year={1989},
  publisher={Elsevier}
}

@article{gilboa2004utilitarian,
  title={Utilitarian aggregation of beliefs and tastes},
  author={Gilboa, Itzhak and Samet, Dov and Schmeidler, David},
  journal={Journal of Political Economy},
  volume={112},
  number={4},
  pages={932--938},
  year={2004},
  publisher={The University of Chicago Press}
}

@article{gilboa2010objective,
  title={Objective and subjective rationality in a multiple prior model},
  author={Gilboa, Itzhak and Maccheroni, Fabio and Marinacci, Massimo and Schmeidler, David},
  journal={Econometrica},
  volume={78},
  number={2},
  pages={755--770},
  year={2010},
  publisher={Wiley Online Library}
}

@article{ghirardato2002ambiguity,
  title={Ambiguity made precise: A comparative foundation},
  author={Ghirardato, Paolo and Marinacci, Massimo},
  journal={Journal of Economic Theory},
  volume={102},
  number={2},
  pages={251--289},
  year={2002},
  publisher={Elsevier}
}

@article{ghirardato2003subjective,
  title={A subjective spin on roulette wheels},
  author={Ghirardato, Paolo and Maccheroni, Fabio and Marinacci, Massimo and Siniscalchi, Marciano},
  journal={Econometrica},
  volume={71},
  number={6},
  pages={1897--1908},
  year={2003},
  publisher={Wiley Online Library}
}

@article{gajdosmaurin2004unequal,
  title={Unequal uncertainties and uncertain inequalities: An axiomatic approach},
  author={Gajdos, Thibault and Maurin, Eric},
  journal={Journal of Economic Theory},
  volume={116},
  number={1},
  pages={93--118},
  year={2004},
  publisher={Elsevier}
}

@article{herstein1953axiomatic,
  title={An axiomatic approach to measurable utility},
  author={Herstein, Israel N. and Milnor, John},
  journal={Econometrica},
  pages={291--297},
  year={1953},
  volume={21},
  number={2},
  publisher={JSTOR}
}

@article{harsanyi1955,
  title={Cardinal welfare, individualistic ethics, and interpersonal comparisons of utility},
  author={Harsanyi, John C.},
  journal={Journal of Political Economy},
  volume={63},
  number={4},
  pages={309--321},
  year={1955},
  publisher={The University of Chicago Press}
}

@article{hylland1979impossibility,
  title={The impossibility of {B}ayesian group decision making with separate aggregation of beliefs and values},
  author={Hylland, Aanund and Zeckhauser, Richard},
  journal={Econometrica},
  pages={1321--1336},
  year={1979},
  volume={47},
  number={6},
  publisher={JSTOR}
}

@article{karni1998impartiality,
  title={Impartiality: Definition and representation},
  author={Karni, Edi},
  journal={Econometrica},
  volume={66},
  number={6},
  pages={1405--1415},
  year={1998},
  publisher={JSTOR}
}

@article{karni2024impartiality,
  title={Impartiality and relative utilitarianism},
  author={Karni, Edi and Weymark, John A.},
  journal={Social Choice and Welfare},
  volume={63},
  number={1},
  pages={1--18},
  year={2024},
  publisher={Springer}
}

@article{kalai1975other,
  title={Other solutions to {N}ash's bargaining problem},
  author={Kalai, Ehud and Smorodinsky, Meir},
  journal={Econometrica},
  volume={43},
  number={3},
  pages={513--518},
  year={1975},
  publisher={Econometric Society}
}

@techreport{kurata2026collective, 
    title={Collective Decisions under Uncertainty: Efficiency, Ex-ante Fairness, and Normalization}, 
    author={Leo Kurata and Kensei Nakamura},
    institution={Graduate School of Economics, Hitotsubashi University},
    series= {Working Papers},
    type= {Discussion paper},
    number={2026–01},
    year={2026}, 
    month={Jan},
}

@article{maccheroni2006ambiguity,
  title={Ambiguity aversion, robustness, and the variational representation of preferences},
  author={Maccheroni, Fabio and Marinacci, Massimo and Rustichini, Aldo},
  journal={Econometrica},
  volume={74},
  number={6},
  pages={1447--1498},
  year={2006},
  publisher={Wiley Online Library}
}

@Article{marchant2018wp,
  author={Thierry Marchant},
  title={{Utilitarianism without individual utilities}},
  journal={Social Choice and Welfare},
  year=2019,
  volume={53},
  number={1},
  pages={1-19},
  month={June},
  doi={},
  url={}
}

@article{mongin1995consistent,
  title={Consistent {B}ayesian aggregation},
  author={Mongin, Philippe},
  journal={Journal of Economic Theory},
  volume={66},
  number={2},
  pages={313--351},
  year={1995},
  publisher={Elsevier}
}

@article{mongin1998paradox,
  title={The paradox of the {B}ayesian experts and state-dependent utility theory},
  author={Mongin, Philippe},
  journal={Journal of Mathematical Economics},
  volume={29},
  number={3},
  pages={331--361},
  year={1998},
  publisher={Elsevier}
}

@article{mongin2021rawls,
  title={Rawls's difference principle and maximin rule of allocation: A new analysis},
  author={Mongin, Philippe and Pivato, Marcus},
  journal={Economic Theory},
  volume={71},
  number={4},
  pages={1499--1525},
  year={2021},
  publisher={Springer}
}

@article{nash1950bargaining,
  title={The bargaining problem},
  author={Nash, John F.},
  journal={Econometrica},
  volume={18},
  number={2},
  pages={155--162},
  year={1950},
  publisher={Econometric Society}
}

@article{nakamura2026social,
  title={Social choice rules when individuals are responsible for their skills},
  author={Kensei Nakamura},
  journal={Journal of Mathematical Economics},
  volume={126},
  pages={103295},
  year={2026},
  publisher={Elsevier}
}

@article{pivato2009twofold,
  title={Twofold optimality of the relative utilitarian bargaining solution},
  author={Pivato, Marcus},
  journal={Social Choice and Welfare},
  volume={32},
  number={1},
  pages={79--92},
  year={2009},
  publisher={Springer}
}

@book{rawls1971,
  author = {Rawls, John},
  publisher = {Cambridge Massachusetts: Harvard University Press, and Oxford: Clarendon Press},
  title = {A Theory of Justice},
  year = 1971
}

@book{savage1,
  title={The Foundations of Statistics},
  author={Savage, Leonard J.},
  year={1954},
  publisher={New York: Wiley}
}

@article{segal2000let,
  title={Let's agree that all dictatorships are equally bad},
  author={Segal, Uzi},
  journal={Journal of Political Economy},
  volume={108},
  number={3},
  pages={569--589},
  year={2000},
  publisher={The University of Chicago Press}
}

@article{sen1976welfare,
  title={Welfare inequalities and {R}awlsian axiomatics},
  author={Sen, Amartya},
  journal={Theory and Decision},
  volume={7},
  number={4},
  pages={243--262},
  year={1976},
  publisher={Springer}
}

@article{sprumont2013relative,
  title={On relative egalitarianism},
  author={Sprumont, Yves},
  journal={Social Choice and Welfare},
  volume={40},
  number={4},
  pages={1015--1032},
  year={2013},
  publisher={Springer}
}

@article{sprumont2019relative,
  title={Relative utilitarianism under uncertainty},
  author={Sprumont, Yves},
  journal={Social Choice and Welfare},
  volume={53},
  number={4},
  pages={621--639},
  year={2019},
  publisher={Springer}
}

@article{sprumont2018belief,
  title={Belief-weighted {N}ash aggregation of {S}avage preferences},
  author={Sprumont, Yves},
  journal={Journal of Economic Theory},
  volume={178},
  pages={222--245},
  year={2018},
  publisher={Elsevier}
}

@article{sprumont2025two,
  title={Two time-consistent {P}aretian solutions to the intertemporal resource allocation problem},
  author={Sprumont, Yves},
  journal={Journal of Economic Theory},
  pages={106048},
  volume={228},
  year={2025},
  publisher={Elsevier}
}

@article{stanca2021smooth,
  title={Smooth aggregation of {B}ayesian experts},
  author={Stanca, Lorenzo},
  journal={Journal of Economic Theory},
  volume={196},
  pages={105308},
  year={2021},
  publisher={Elsevier}
}

@incollection{weymark1991, 
    place={Cambridge}, 
    title={A reconsideration of the {H}arsanyi--{S}en debate on utilitarianism}, 
    booktitle={Interpersonal Comparisons of Well-Being}, 
    publisher={Cambridge University Press}, 
    author={Weymark, John A.}, 
    editor={Elster, Jon and Roemer, John E.}, 
    year={1991}, 
    pages={255--320}
    }

\end{document}